\documentclass[10pt]{article}

\usepackage[a4paper,margin=1in]{geometry}
\usepackage{jheppub}
\usepackage[T1]{fontenc}
\usepackage{graphicx}
\usepackage{amssymb}
\usepackage{amsmath}
\usepackage{dsfont}
\usepackage{setspace}
\usepackage{slashed}
\usepackage{pifont}
\newcommand{\cmark}{\ding{51}}%
\newcommand{\xmark}{\ding{55}}%
\usepackage[capitalise]{cleveref}
\usepackage{multirow}
\usepackage{makecell}
\usepackage{tabularx}
\usepackage{comment}
\newcolumntype{C}[1]{>{\hsize=#1\hsize\centering\arraybackslash}X}%
\usepackage{subcaption}

\newcolumntype{Z}{r<{\hspace{3mm}}}

\renewcommand{\arraystretch}{1.4}
\renewcommand{\i}{\ensuremath{\mathrm{i}}}

\allowdisplaybreaks

\def\lc{\mathrm{lc}}

\def\cH{\mathcal{H}}
\def\cO{\mathcal{O}}

\def\eps{\epsilon}
\DeclareMathOperator{\tr}{\rm tr}

\def\trm{\tr_-}
\def\trp{\tr_+}

\def\colhelsum{\underset{\mathrm{colour}}{\overline{\sum}}\,\underset{\mathrm{helicity}}{\overline{\sum}}}

\def\GeV{\mathrm{GeV}}

\def\waa{W\gamma\gamma}
\def\wpaa{W^+\gamma\gamma}
\def\wmaa{W^-\gamma\gamma}

\def\la{\langle}
\def\ra{\rangle}
\def\spA#1#2{\la#1#2\ra}
\def\spB#1#2{[#1#2]}
\def\spAB#1#2#3{\la#1|#2|#3]}

\def\spAA#1#2#3{\la#1|#2|#3\ra}

\def\colhelsum{\underset{\mathrm{colour}}{\overline{\sum}}\,\underset{\mathrm{helicity}}{\overline{\sum}}}

\allowdisplaybreaks

\title{Two-loop amplitudes for $\cO(\alpha_s^2)$ corrections to $\waa$ production at the LHC}

\author[a]{Simon Badger,}
\author[b,c]{Heribertus Bayu Hartanto,}
\author[d]{Zihao Wu,}
\author[e,f]{Yang Zhang,}
\author[g]{Simone Zoia}

\affiliation[a]{Dipartimento di Fisica and Arnold-Regge Center, Università di Torino, and INFN, Sezione di Torino, Via P.\ Giuria 1, I-10125 Torino, Italy}
\affiliation[b]{Asia Pacific Center for Theoretical Physics, Pohang, 37673, Korea}
\affiliation[c]{Department of Physics, Pohang University of Science and Technology, Pohang, 37673, Korea}
\affiliation[d]{School of Fundamental Physics and Mathematical Sciences, Hangzhou Institute for Advanced Study, UCAS, Hangzhou, 310000, China}
\affiliation[e]{Interdisciplinary Center for Theoretical Study, University of Science and Technology of China, Hefei, Anhui 230026, China}
\affiliation[f]{Peng Huanwu Center for Fundamental Theory, Hefei, Anhui 230026, China}
\affiliation[g]{CERN, Theoretical Physics Department, CH-1211 Geneva 23, Switzerland}

\emailAdd{simondavid.badger@unito.it}
\emailAdd{bayu.hartanto@apctp.org}
\emailAdd{wuzihao@mail.ustc.edu.cn}
\emailAdd{yzhphy@ustc.edu.cn}
\emailAdd{simone.zoia@cern.ch}

\preprint{
\begin{tabular}{l}
   CERN-TH-2024-146 \\
   USTC-ICTS/PCFT-24-29
\end{tabular}
}

\abstract{
We present the two-loop helicity amplitudes contributing to the next-to-next-to-leading order QCD predictions for $W$-boson production in association with two photons at the Large Hadron Collider.
We derived compact analytic expressions for the two-loop amplitudes in the leading colour limit, and 
provide numerical results for the subleading colour contributions.
We employ a compact system of integration-by-part identities provided by the \textsc{NeatIBP} package,
allowing for an efficient computation of the rational coefficients of the scattering amplitudes over finite fields.
}

\begin{document}
\maketitle
\flushbottom

\section{Introduction}

Probing the electroweak sector of the Standard Model of Particle Physics~(SM) can provide us with a deeper understanding 
of the underlying mechanism of spontaneous symmetry breaking, and at the same time offers a window to investigate the presence of new physics beyond the Standard Model~(BSM). 
In particular, multi vector boson production at the Large Hadron Collider~(LHC) is a class of processes that can be utilised to 
study the vector boson self couplings~\cite{Green:2016trm}. 
Triple vector boson production, for instance, can be used to study triple and quartic vector boson couplings simultaneously. 
The production of a $W$ boson in association with two photons ($\waa$ production) is one of the multi vector boson 
processes that can be employed to study the $W$-boson couplings to photons ($WW\gamma$ and $WW\gamma\gamma$ 
interactions)~\cite{Eboli:2000ad,Bell:2009vh}. 
Hints for BSM mechanisms can be searched for by looking for deviations of the measured triple and quartic vector-boson couplings from the SM prediction. 
Such deviations can be probed systematically using the framework of Standard Model Effective Field Theory 
(SMEFT)~\cite{Degrande:2012wf,Falkowski:2016cxu,Celada:2024cxw}.

$\waa$ production has been observed at the LHC, with cross section measurements performed by the ATLAS~\cite{ATLAS:2015ify,ATLAS:2023avk} and CMS~\cite{CMS:2017tzy,CMS:2021jji} experiments.
Theoretical predictions for $\waa$ production  are known at leading order~(LO)~\cite{Baur:1999hm,Baur:1997bn} and 
next-to-leading order~(NLO) accuracy, including both Quantum Chromodynamics~(QCD)~\cite{Baur:2010zf,Bozzi:2011wwa} and electroweak~(EW)~\cite{Greiner:2017mft} corrections.
Analogously to $W\gamma$ production, $\waa$ production features a radiation-amplitude zero (a situation where the amplitude vanishes exactly for a certain phase-space configuration) at LO, 
although it is lifted by higher-order corrections~\cite{Baur:1997bn}. 
The NLO QCD corrections are found to be very large, reaching $\approx~250\%$~\cite{Bozzi:2011wwa,Kim:2024ppt}, due to gluon-initiated channels that open up at this order.
The NLO QCD corrections were computed also for $pp\to\waa +j$~\cite{Campanario:2011ud}, in order to study the impact of higher-order corrections to the theoretical uncertainty due to renormalisation-scale variation on the $W\gamma\gamma j$ contribution to $W\gamma\gamma$
and to capture part of the next-to-next-to leading order~(NNLO) QCD corrections to the inclusive $\waa$ cross section.
It is therefore important for a full-fledged NNLO QCD calculation of $\waa$ production to be performed in order to assess 
the perturbative convergence of the theoretical predictions.

One of the main bottlenecks towards obtaining NNLO QCD prediction for $\waa$ production is the computation of the required two-loop 
QCD corrections to a five-particle scattering process with an off-shell leg. 
Both for the massless and one-external-mass cases, the full set of two-loop five-point master integrals have been 
computed~\cite{Gehrmann:2015bfy,Chicherin:2018old,Abreu:2018rcw,Chicherin:2018mue,Abreu:2020jxa,Canko:2020ylt,Abreu:2021smk,Kardos:2022tpo,Abreu:2023rco}.
They are expressed in terms of \textit{pentagon functions}~\cite{Gehrmann:2018yef,Chicherin:2020oor,Chicherin:2021dyp,Abreu:2023rco}, which allow for an efficient numerical evaluation by means of a public library~\cite{PentagonFunctions:cpp}.
By now, all the massless two-loop QCD five-point amplitudes relevant for LHC physics are known analytically in both the leading colour 
approximation~\cite{Gehrmann:2015bfy,Badger:2018enw,Abreu:2018zmy,Abreu:2019odu,Abreu:2020cwb,Chawdhry:2020for,Agarwal:2021grm,Chawdhry:2021mkw,Abreu:2021oya} 
and full colour~\cite{Badger:2019djh,Agarwal:2021vdh,Badger:2021imn,Abreu:2023bdp,Badger:2023mgf,Agarwal:2023suw,DeLaurentis:2023nss,DeLaurentis:2023izi}.
For the two-loop five-point amplitudes with an external massive
leg, instead, analytic expressions have been obtained only in the leading colour approximation, 
more specifically by taking into account only the contributions from the planar diagrams~\cite{Badger:2021nhg,Badger:2021ega,Abreu:2021asb,Badger:2022ncb}.
Although the non-planar master integrals are already available, 
the subleading colour amplitudes remain challenging to obtain due to high algebraic complexity of the rational functions multiplying
the master integrals or the pentagon functions.
Nevertheless, the availability of the two-loop scattering amplitudes, even in the leading colour approximation, have made it possible
for several $2\to 3$ theoretical predictions to be computed at NNLO QCD 
accuracy~\cite{Chawdhry:2019bji,Kallweit:2020gcp,Chawdhry:2021hkp,Czakon:2021mjy,Badger:2021ohm,
Alvarez:2023fhi,Badger:2023mgf,Hartanto:2022qhh,Hartanto:2022ypo,Buonocore:2022pqq,Catani:2022mfv,Buonocore:2023ljm,Mazzitelli:2024ura}. 
Efforts to include more massive particles or more external legs have been recently made as well~\cite{Henn:2021cyv,Badger:2022hno,FebresCordero:2023gjh,Agarwal:2024jyq,Henn:2024ngj,Badger:2024fgb,Jiang:2024eaj,Abreu:2024flk,Chestnov:2024mnw}.

In this work, we tackle the two-loop scattering amplitude for $\waa$ production at the LHC. We obtain an analytic expression of the 
two-loop amplitude in the leading colour approximation, including the contributions from the non-planar Feynman diagrams. 
We employ the state-of-the-art methodology based on functional reconstruction from numerical evaluations over finite fields to overcome the algebraic complexity~\cite{vonManteuffel:2014ixa,Peraro:2016wsq,Klappert:2019emp,Peraro:2019svx,Klappert:2020aqs,Klappert:2020nbg}. 
In addition, we benefit from compact systems of integration-by-parts~(IBP) relations obtained with \textsc{NeatIBP}~\cite{Wu:2023upw}.
The analytic reconstruction of the subleading colour amplitudes, however,
is extremely expensive in comparison to the leading colour ones. 
As an alternative to the analytic computation, we take a numerical approach, still
within the finite-field framework: we reconstruct the \emph{values} of the rational coefficients of 
the pentagon-function monomials at rationalised 
physical phase-space points from their evaluations over several prime fields.
While this approach is substantially more expensive than the evaluation of an analytic expression for the amplitude, 
it may be nonetheless viable in those scenarios where the contribution from the two-loop amplitudes is small such that the
subleading colour amplitudes can be evaluated at a relatively small set of phase-space points.
Furthermore, these results, although partially numerical, allow us to study at amplitude level the non-trivial analytic properties of the non-planar Feynman integrals observed in Ref.~\cite{Abreu:2023rco}.

This paper is organised as follows. In section~\ref{sec:amplitude} we present the general structure of the $\waa$ amplitude up to two loops, including its decomposition into five-, four- and three-particle $W$-production amplitudes with
their corresponding decay currents. 
In section~\ref{sec:framework} we discuss the finite-field computational framework that we apply to compute helicity amplitudes in both the analytical and 
the numerical approaches. We present numerical benchmark results, discuss the analytic properties of the amplitude
and describe the ancillary files containing analytic expressions of the leading colour amplitude in section~\ref{sec:result}. 
We finally draw our conclusions and outline potential future developments in section~\ref{sec:conclusions}.

\section{Structure of the $\waa$ amplitude}
\label{sec:amplitude}

We consider the production of a $W$ boson in association with two photons at the LHC ($W\gamma \gamma$), with decay of the $W$ boson to a leptonic pair.
Without any loss of generality we focus on the production of a $W^+$.
The amplitude for $W^-\gamma \gamma$ can then be obtained from the one for $W^+\gamma \gamma$ through appropriate permutations and parity conjugations, as explained below.
There is only one partonic channel, initiated by a quark ($d$) and an anti-quark ($\bar{u}$).
We assign momenta $p_i$ and helicity states $h_i$ as
\begin{equation}
0 \rightarrow \bar{u}(p_1,h_1)+d(p_2,h_2)+\gamma(p_3,h_3)+\gamma(p_4,h_4)+\nu_\ell(p_5,h_5)+\ell^+(p_6,h_6) \,.
\label{eq:processdefinition}
\end{equation}
All the external momenta are taken to be outgoing and satisfy the on-shell massless conditions,
\begin{equation}
\sum_{i=1}^{6} p_i = 0 \,, \qquad \qquad p_i^2 = 0\quad \forall \; i = 1, \dots,6 \,.
\label{eq:momentumconservation}
\end{equation}
We work in dimensional regularisation and present our results in the ’t~Hooft-Veltman (tHV) scheme, with $d=4-2\eps$ space-time dimensions and four-dimensional external momenta.
The fact that the external momenta are four-dimensional implies a further constraint on the kinematics, which can be expressed as the vanishing of the Gram determinant of the five independent external momenta (see Appendix~\ref{app:rationalise6pt}).
This reduces the number of independent kinematic variables from $9$ to $8$.

In order to achieve a minimal and rational parametrisation of the kinematics, we employ a momentum-twistor parametrisation~\cite{Hodges:2009hk,Badger:2016uuq}.
In particular, we make use of the parametrisation proposed in Ref.~\cite{Badger:2021ega}.
We denote the six-particle momentum-twistor variables by $\vec{z} = (z_1,\ldots,z_8)$, and report their definition in Appendix~\ref{app:MomentumTwistors}.
We emphasise that, in addition to providing a minimal parametrisation of the kinematics, momentum-twistor variables also allow us to express the spinor-helicity variables as rational functions.

\subsection{Amplitude decomposition}
\label{sec:Decomposition}

\begin{figure}[t!]
  \begin{center}
    \includegraphics[width=0.75\textwidth]{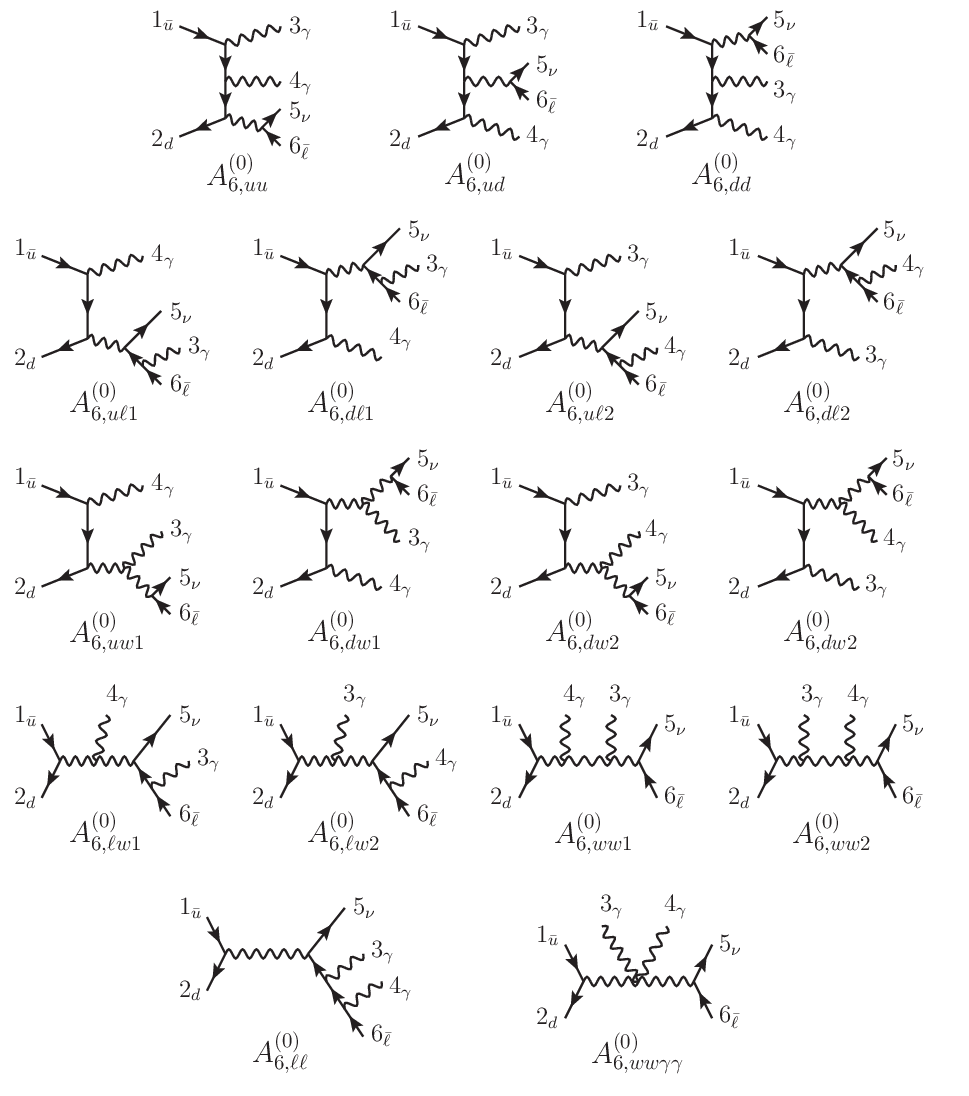}
  \end{center}
	\caption{Representative tree-level Feynman diagrams for the sub-amplitudes appearing in Eq.~\eqref{eq:ampdecomposition}. 
	$A^{(L)}_{6,q}$ starts to contribute only at the two loop level.}
  \label{fig:diag0L}
\end{figure}

We construct the partial amplitude $A_{6}^{(L)}$ for $W^+(\to \nu_{\ell} \ell^+)\gamma \gamma$ production by stripping off 
the colour and loop factors as well as the coupling constants from the full amplitude $M_{6}^{(L)}$ as
\begin{equation}
M_{6}^{(L)}(1_{\bar{u}},2_d,3_\gamma,4_\gamma,5_{\nu},6_{\bar{\ell}}) = e^2 g_W^2 \, 
	\bigg[ (4\pi)^\eps e^{-\eps \gamma_E} \frac{\alpha_s}{4\pi} \bigg]^L \, 
	\delta_{i_2}^{\;\;\bar i_1} \, 
	A_{6}^{(L)}(1_{\bar{u}},2_d,3_\gamma,4_\gamma,5_{\nu},6_{\bar{\ell}}) \,. 
\label{eq:colourdecomposition}
\end{equation}
Here, $e$ is the electric charge, $g_W$ is the weak coupling constant, $\eps$ is the dimensional regularisation parameter ($\eps=(4-d)/2$),
and $\delta_{i_2}^{\;\;\bar i_1}$ is the colour factor. 
The final state photons can be emitted either by the
initial state quarks, by the $W$ boson or by the charged lepton. We can therefore decompose the partial amplitude according to the source of photon radiation, as follows: 
\begin{align}
\label{eq:ampdecomposition}
	\begin{aligned}
	A_{6}^{(L)}   = &\bigg[  Q_u^2 A^{(L)}_{6,uu} + Q_u Q_d A^{(L)}_{6,ud} + Q_d^2 A^{(L)}_{6,dd} + \bigg(\sum_{q=1}^{n_f} Q_q^2 \bigg) A^{(L)}_{6,q} \bigg] P(s_{56})  \\
                & + \bigg[ Q_u Q_\ell A^{(L)}_{6,u\ell 1} + Q_d Q_\ell A^{(L)}_{6,d\ell 1}  \bigg] P(s_{356}) 
                  + \bigg[ Q_u Q_\ell A^{(L)}_{6,u\ell 2} + Q_d Q_\ell A^{(L)}_{6,d\ell 2}  \bigg] P(s_{456})  \\
                & + \bigg[ Q_u Q_w A^{(L)}_{6,uw1} + Q_d Q_w A^{(L)}_{6,dw1}   \bigg] P(s_{356})P(s_{56})  \\
                & + \bigg[ Q_u Q_w A^{(L)}_{6,uw2} + Q_d Q_w A^{(L)}_{6,dw2}   \bigg] P(s_{456})P(s_{56})  \\
                & + Q_\ell Q_w A^{(L)}_{6,\ell w1}P(s_{356})P(s_{3456}) + Q_\ell Q_w A^{(L)}_{6,\ell w2}P(s_{456})P(s_{3456})  \\
                & + Q_w^2 A^{(L)}_{6,ww1}P(s_{56})P(s_{356})P(s_{3456}) + Q_w^2 A^{(L)}_{6,ww2}P(s_{56})P(s_{456})P(s_{3456})  \\
                & + Q_\ell^2 A^{(L)}_{6,\ell\ell}P(s_{3456}) +  A^{(L)}_{6,ww\gamma\gamma}P(s_{56})P(s_{3456}) \,,
	\end{aligned}
\end{align}
where
\begin{equation}
P(s) = \frac{1}{s-\mu_W^2}
\end{equation}
is the denominator factor of $W$-boson propagator, $s_{i\dots k}$ are the Mandelstam invariants,
\begin{equation}
s_{i\dots k} = (p_i+ \cdots + p_k)^2 \,,
\end{equation} 
$Q_i$ is the fractional electric charge of the $i$-th particle, $\mu_W$ is the mass of $W$ boson
and $n_f$ is the number of light quarks. 
In Fig.~\ref{fig:diag0L} we show representative tree-level Feynman diagrams for each sub-amplitude appearing in Eq.~\eqref{eq:ampdecomposition}.
The sub-amplitude where both photons are attached to the internal quark loop, $A^{(L)}_{6,q}$, vanishes at tree level and one loop,
and starts to contribute only at the two-loop level. 
We also note that the $A^{(2)}_{6,q}$ amplitude contains two-loop non-planar Feynman diagrams in the leading colour limit.
Furthermore, the sub-amplitudes $A^{(L)}_{6,i}$ in Eq.~\eqref{eq:ampdecomposition} are 
not separately gauge invariant in the electroweak sector except for $A^{(L)}_{6,q}$.

The loop-level sub-amplitudes can be further decomposed according to the number of colours~($N_c$) and closed light quark loops~($n_f$), as
\begin{align}
	A^{(1)}_{6,i} & = \bigg( N_c -\frac{1}{N_c} \bigg)  A^{(1),N_c}_{6,i} \,, \label{eq:NcNfdecomposition1L}  \\
	A^{(2)}_{6,i} & = N_c^2 A^{(2),N_c^2}_{6,i} -  \left(  A^{(2),N_c^2}_{6,i} + A^{(2),1/N_c^2}_{6,i} \right) + \frac{1}{N_c^2} A^{(2),1/N_c^2}_{6,i} 
			  + \bigg(N_c-\frac{1}{N_c}\bigg) n_f A^{(2),N_c n_f}_{6,i}  \,, \label{eq:NcNfdecomposition2L} \\
	A^{(2)}_{6,q} & =  \bigg(N_c-\frac{1}{N_c}\bigg) A^{(2),N_c}_{6,q} \,. \label{eq:NcNfdecomposition2Lqq}
\end{align}
Here, $A^{(L)}_{6,i}$ are the sub-amplitudes appearing in Eq.~\eqref{eq:ampdecomposition}
where neither of the photons are coupled to the closed quark loop.
In the leading colour approximation, the one- and two-loop amplitude decompositions become
\begin{align}
	A^{(1),\lc}_{6,i} & = N_c A^{(1),N_c}_{6,i} \,, \label{eq:NcNfdecomposition1Llc} \\
	A^{(2),\lc}_{6,i} & = N_c^2 A^{(2),N_c^2}_{6,i} +  N_c n_f A^{(2),N_c n_f}_{6,i} \,, \label{eq:NcNfdecomposition2Llc} \\
	A^{(2),\lc}_{6,q} & = N_c A^{(2),N_c}_{6,q} \,. \label{eq:NcNfdecomposition2Lqqlc}
\end{align}

In our helicity-amplitude computation, the V--A coupling of the $W$ boson to fermions selects a subset of helicity configurations
contributing to the amplitude. They are
\begin{align}
\begin{aligned}
\label{eq:sixpthelicityamps}
& A_{6}^{(L)}(1^+_{\bar{u}},2^-_d,3^+_\gamma,4^+_\gamma,5^-_{\nu},6^+_{\bar{\ell}}) \,, \\
& A_{6}^{(L)}(1^+_{\bar{u}},2^-_d,3^+_\gamma,4^-_\gamma,5^-_{\nu},6^+_{\bar{\ell}}) \,, \\
& A_{6}^{(L)}(1^+_{\bar{u}},2^-_d,3^-_\gamma,4^+_\gamma,5^-_{\nu},6^+_{\bar{\ell}}) \,, \\
& A_{6}^{(L)}(1^+_{\bar{u}},2^-_d,3^-_\gamma,4^-_\gamma,5^-_{\nu},6^+_{\bar{\ell}}) \,.
\end{aligned}
\end{align}

We define the complex-conjugated process, $W^-(\to \bar{\nu}_{\ell} \ell^- ) \gamma \gamma$ production, as
\begin{equation}
0 \rightarrow \bar{d}(p_1,h_1)+u(p_2,h_2)+\gamma(p_3,h_3)+\gamma(p_4,h_4)+\ell^-(p_5,h_5)+\bar{\nu}_\ell(p_6,h_6) \,.
\end{equation} 
The $\wmaa$ helicity amplitudes can be obtained from the $\wpaa$ ones by exchanging the momenta of the quark-anti quark pair 
as well as of the lepton-neutrino pair, followed by parity conjugation:
\begin{equation}
	A_{6}^{(L)}(1^{+}_{\bar{d}},2^{-}_u,3^{h_3}_\gamma,4^{h_4}_\gamma,5^{-}_{\ell},6^{+}_{\bar{\nu}}) = 
    \left. A_{6}^{(L)}(2^{+}_{\bar{u}},1^{-}_d,3^{-h_3}_\gamma,4^{-h_4}_\gamma,6^{-}_{\nu},5^{+}_{\bar{\ell}})\right|_{\spA{i}{j} 
\leftrightarrow \spB{j}{i}} \,.
\end{equation}

\subsection{Finite remainder}

We construct the finite remainders $F^{(L)}_{6,i}$ by subtracting the ultraviolet~(UV) and infrared~(IR) singularities 
from the bare loop sub-amplitudes $A^{(L)}_{6,i}$,
\begin{equation}
	F^{(L)}_{6,i} = \underset{\eps \to 0}{\lim} \left[ A^{(L)}_{6,i} - \mathcal{P}^{(L)}_{6,i}\right] \,. 
    \label{eq:finrem}
\end{equation}
The pole terms $\mathcal{P}^{(L)}_{6,i}$ are built up of the UV counter-terms (where the $\alpha_s$ is renormalised in the $\overline{\rm MS}$ scheme)
and known universal IR poles~\cite{Catani:1998bh,Becher:2009qa,Becher:2009cu,Gardi:2009qi},
\begin{align}
	\mathcal{P}^{(1)}_{6,i} & = 2 \, I_1(\eps) A^{(0)}_{6,i} \,, \label{eq:pole1L} \\
	\mathcal{P}^{(2)}_{6,i} & = 4 \, I_2(\eps) A^{(0)}_{6,i} + 2 \bigg( I_1(\eps) + \frac{\beta_0}{\eps} \bigg) A^{(1)}_{6,i}  \,, \label{eq:pole2L}
\end{align}
where $I_1(\eps)$ and $I_2(\eps)$ are the `Catani operators', given by~\cite{Gehrmann:2011ab}
\begin{align}
	I_1(\eps) & = -\frac{N(\eps)}{2} \bigg(\frac{N_c}{2} - \frac{1}{2N_c} \bigg) \bigg[ \frac{2}{\eps^2} + \frac{3}{\eps} \bigg] \left(-\frac{s_{23}+\i 0^+}{\mu_R}\right)^{-\eps} \,,
    \label{eq:I1operator} \\
	I_2(\eps) & = -\frac{1}{2} I_1(\eps) \bigg[ I_1(\eps) - \frac{\beta_0}{\eps}\bigg]
  	              + \frac{N(\eps)}{N(2\eps)} \bigg[ \frac{\beta_0}{\eps} + K \bigg] I_1(2\eps) + H^{(2)}(\eps) \,,
    \label{eq:I2operator}
\end{align}
where $0^+$ is a positive infinitesimal, with
\begin{align}
	H^{(2)}(\eps) & = \frac{N(\eps)}{2} \left( N_c^2 -1 \right) 
	\bigg[ \frac{7\zeta_3}{4} + \frac{409}{864} - \frac{11\pi^2}{96}
	      +\frac{1}{N_c^2}\bigg( \frac{3\zeta_3}{2} + \frac{3}{32} - \frac{\pi^2}{8} \bigg) \nonumber \\
	      & \qquad\qquad\qquad\qquad\quad + \frac{n_f}{N_c}\bigg( \frac{\pi^2}{48} - \frac{25}{216} \bigg) \bigg] \,, \\
	K & = \frac{67}{18} - \frac{\pi^2}{6} C_A - \frac{10}{9} T_R n_f \,, \\
	N(\eps) & = \frac{e^{\eps \gamma_E}}{\Gamma(1-\eps)} \,.
\end{align} 
The coefficient of the $\beta$ function are~\cite{Catani:1998bh,Becher:2009qa}
\begin{align}
\beta_0 & = \frac{11}{6}C_A - \frac{2}{3} T_R n_f \,, \\
\beta_1 & = \frac{17}{6} C_A^2 - \frac{5}{3}C_A T_R n_f - C_F T_R n_f  \,,
\end{align}
where
\begin{align}
T_R = \frac{1}{2}  \,, \qquad \qquad C_A = N_c \,, \qquad \qquad C_F = \frac{N_c^2-1}{2 N_c} \,.
\end{align}

We calculated the finite remainders setting the renormalisation scale ($\mu_R$) to one.
We then derived the $\mu_R$ dependence of the finite remainders directly from Eqs.~\eqref{eq:finrem}~--~\eqref{eq:pole2L}
by restoring the $\mu_R$ dependence in both the Catani operators ($I_1(\eps)$ and $I_2(\eps)$ in Eqs.~\eqref{eq:I1operator}~and~\eqref{eq:I2operator}) and the renormalisation counterterms.
We obtain
\begin{equation}
	F_{6,i}^{(L)}(\mu_R^2) = F_{6,i}^{(L)}(\mu_R^2 = 1) + \delta F_{6,i}^{(L)}(\mu_R^2) \,,
	\label{eq:mudepF}
\end{equation}
where
\begin{align}
	\delta F_{6,i}^{(L),N_c^2}(\mu_R^2) & = \log(\mu_R^2) \bigg\lbrace \bigg( \frac{409}{216} 
	  - \frac{11}{24} \pi^2 + 7\zeta_3 \bigg) A_{6,i}^{(0)} + \frac{11}{3} F_{6,i}^{(1),N_c}(1) \bigg\rbrace \,, \label{eq:mudepNcp2} \\
	\delta F_{6,i}^{(L),1/N_c^2}(\mu_R^2) & = \log(\mu_R^2) \bigg\lbrace -\frac{3}{8} 
	  + \frac{\pi^2}{2}  - 6\zeta_3 \bigg\rbrace A_{6,i}^{(0)}  \,,  \label{eq:mudepNcpm2} \\
	\delta F_{6,i}^{(L),N_c n_f}(\mu_R^2) & = \log(\mu_R^2) \bigg\lbrace \bigg( -\frac{25}{54} 
	  + \frac{\pi^2}{12}  \bigg) A_{6,i}^{(0)} - \frac{2}{3} F_{6,i}^{(1),N_c}(1) \bigg\rbrace \,.   \label{eq:mudepNcNf} 
\end{align}
The above $\mu_R$-restoring terms have been split up according to the $(N_c, n_f)$ decomposition as specified in Eqs.~\eqref{eq:NcNfdecomposition1L}~--~\eqref{eq:NcNfdecomposition2Lqq}.
The remaining $(N_c, n_f)$-components which are not listed in Eqs.~\eqref{eq:mudepNcp2}~--~\eqref{eq:mudepNcNf} do not depend on $\mu_R$.

The hard functions, obtained by squaring the finite remainders and averaging over all colour and helicity states, are defined by 
\begin{align}
	\cH^{(0)} &= e^4 g_W^4 N_c \; \colhelsum F^{(0)*}_{6} F^{(0)}_{6} \,,  \label{eq:hardfunction0L}\\
	\cH^{(1)} &= e^4 g_W^4 N_c \bigg(\frac{\alpha_s}{4\pi}\bigg)   \; \colhelsum 2 F^{(0)*}_{6} F^{(1)}_{6} \,, \label{eq:hardfunction1L} \\
	\cH^{(2)} &= e^4 g_W^4 N_c \bigg(\frac{\alpha_s}{4\pi}\bigg)^2 \; \colhelsum \bigg[ 2 F^{(0)*}_{6} F^{(2)}_{6} + F^{(1)*}_{6} F^{(1)}_{6}\bigg] \,,
     \label{eq:hardfunction2L}
\end{align}
where the bar over the sum sign indicates the average. The loop-level hard functions depend on the $\mu_R$ through the finite remainders (c.f.~Eq.~\ref{eq:mudepF}) 
and the running strong coupling constant $\alpha_s(\mu_R)$.

\section{Helicity-amplitude computation}
\label{sec:framework}

In this section we describe our strategy to compute the helicity amplitudes for $\waa$ production up to two loops.
We begin by identifying the independent building blocks that make up the six-point amplitude.
Next, we present our computational framework, first for the analytic calculation at leading colour and then for the numerical calculation at subleading colour. 
We conclude this section by discussing a number of checks.

\subsection{Independent sub-amplitudes}

As we discussed in Section~\ref{sec:Decomposition}, the final state photons in $\waa$ production can be emitted from different sources. 
This is reflected in the decomposition of the amplitude given in Eq.~\eqref{eq:ampdecomposition}. By exploiting the fact that the QCD corrections only
affect the $W$-production amplitude, we can remove the $W$-boson decay current from the six-particle amplitude, and identify the independent lower-multiplicity 
building blocks that need to be computed. 
The full six-particle amplitude can then be recovered by re-attaching the appropriate $W$-boson decay current.
We organise these amplitude building blocks in the following classes.

\begin{itemize}
\item \textit{Five-particle amplitudes} \\
This class is comprised of sub-amplitudes where both photons are coupled to either the internal or the external quark line.
In Eq.~\eqref{eq:ampdecomposition} these are:
\begin{equation}
A^{(L)}_{6,uu} \,,\quad 
A^{(L)}_{6,ud} \,,\quad
A^{(L)}_{6,dd} \,,\quad
A^{(L)}_{6,q} \,. 
\label{eq:amp5ptlist}
\end{equation}
In this category, the six-particle sub-amplitude can be written by contracting the five-particle one with the current for the $W\to\ell \nu$ decay
process, $L_\mu(5_{\nu},6_{\bar{\ell}})$, as
\begin{equation}
\label{eq:5ptcurrentdecomposition}
	A^{(L)}_{6,i}(1_{\bar{u}},2_d,3_\gamma,4_\gamma,5_{\nu},6_{\bar{\ell}})  
	= A^{(L)\mu}_{5,i}(1_{\bar{u}},2_d,3_\gamma,4_\gamma,p_{56}) \, L_\mu(5_{\nu},6_{\bar{\ell}}) \,,
\end{equation}
for the sub-amplitudes $A^{(L)}_{6,i}$ listed in Eq.~\eqref{eq:amp5ptlist}.

\item \textit{Four-particle amplitudes} \\
The four-particle sub-amplitudes are those where one photon is coupled to the quark line and the other is 
attached to either the $W$ boson or the charged lepton. 
The sub-amplitudes in Eq.~\eqref{eq:ampdecomposition} that belong to this category are
\begin{eqnarray}
& A^{(L)}_{6,u\ell 1} \,,\quad
  A^{(L)}_{6,d\ell 1} \,,\quad
  A^{(L)}_{6,uw1} \,,\quad
  A^{(L)}_{6,dw1} \,,
  \label{eq:amp4pt1list} \\
& A^{(L)}_{6,u\ell 2} \,,\quad
  A^{(L)}_{6,d\ell 2}\,,\quad
  A^{(L)}_{6,uw2} \,,\quad
  A^{(L)}_{6,dw2} \,. 
  \label{eq:amp4pt2list}
\end{eqnarray}
For the sub-amplitudes $A^{(L)}_{6,i}$ listed in Eq.~\eqref{eq:amp4pt1list},
the photon with momentum $p_3$ is coupled to the external quark line, and the six-particle sub-amplitudes take the form
\begin{equation}
\label{eq:4ptcurrentdecomposition4}
A^{(L)}_{6,i}(1_{\bar{u}},2_d,3_\gamma,4_\gamma,5_{\nu},6_{\bar{\ell}})  
	 = A^{(L)\mu}_{4,u/d}(1_{\bar{u}},2_d,3_\gamma,p_{456}) \, L^{i}_\mu(4_\gamma,5_{\nu},6_{\bar{\ell}}) \,, 
\end{equation}
where $L^{i}_\mu(4_\gamma,5_{\nu},6_{\bar{\ell}})$ is the $W \to \ell\nu\gamma$ radiative decay current.
For the sub-amplitudes $A^{(L)}_{6,i}$ listed in Eq.~\eqref{eq:amp4pt2list}, on the other hand, the photon with momentum $p_4$ is coupled to the external quark line, and the six-particle sub-amplitudes are given by
\begin{equation}
\label{eq:4ptcurrentdecomposition3}
A^{(L)}_{6,i}(1_{\bar{u}},2_d,3_\gamma,4_\gamma,5_{\nu},6_{\bar{\ell}})  
	 = A^{(L)\mu}_{4,u/d}(1_{\bar{u}},2_d,4_\gamma,p_{356}) \, L^{i}_\mu(3_\gamma,5_{\nu},6_{\bar{\ell}}) \,.
\end{equation}

\item \textit{Three-particle amplitudes} \\
This category is made up of sub-amplitudes where neither of the photons are attached to the quark line:
\begin{equation}
A^{(L)}_{6,\ell w1} \,,\quad
A^{(L)}_{6,\ell w2} \,,\quad
A^{(L)}_{6,ww1} \,,\quad
A^{(L)}_{6,ww2} \,,\quad
A^{(L)}_{6,\ell\ell} \,,\quad
A^{(L)}_{6,ww\gamma\gamma} \,. 
\label{eq:amp3ptlist}
\end{equation}
From these we can detach the $W\to \ell\nu\gamma\gamma$ radiative decay current $L^{i}_\mu(3_\gamma,4_\gamma,5_{\nu},6_{\bar{\ell}})$ to obtain the three-particle sub-amplitudes:
\begin{equation}
\label{eq:3ptcurrentdecomposition}
A^{(L)}_{6,i}(1_{\bar{u}},2_d,3_\gamma,4_\gamma,5_{\nu},6_{\bar{\ell}})  
	 = A^{(L)\mu}_{3}(1_{\bar{u}},2_d,p_{3456}) \, L^{i}_\mu(3_\gamma,4_\gamma,5_{\nu},6_{\bar{\ell}}) \,.
\end{equation}
\end{itemize}

We therefore identify the independent sub-amplitudes to be $A^{(L)\mu}_{5,uu/ud}$, $A^{(L)\mu}_{4,u}$ and $A^{(L)\mu}_{3}$.
We then compute these independent five-, four- and three-particle sub-amplitudes by first decomposing them into form factors, as
\begin{align}
	A^{(L)\mu}_{5,uu/ud} & = \sum_{i=1}^{4} \Omega^{(L)}_{uu/ud;i} \; u^{\mu}_{i} \;, \\
	A^{(L)\mu}_{4,u}     & = \sum_{i=1}^{4} \Omega^{(L)}_{u;i} \; v^{\mu}_{i} \;, \\
	A^{(L)\mu}_{3}       & = \sum_{i=1}^{4} \Omega^{(L)}_{3;i} \; w^{\mu}_{i} \;,
\end{align}
where $\{u_i\}$, $\{v_i\}$ and $\{w_i\}$ are the spanning bases for the five-, four- and three-particle amplitudes respectively, given by
\begin{align}
u^\mu_1 & = p^\mu_1 \,, &
u^\mu_2 & = p^\mu_2 \,, & 
u^\mu_3 & = p^\mu_3 \,, & 
u^\mu_4 & = p^\mu_4 \,, \\
v^\mu_1 & = p^\mu_1 \,, & 
v^\mu_2 & = p^\mu_2 \,, & 
v^\mu_3 & = p^\mu_3 \,, & 
v^\mu_4 & = \frac{ \spAB{2}{3}{1} \spAB{1}{\gamma^\mu}{2} 
          - \spAB{1}{3}{2} \spAB{2}{\gamma^\mu}{1} }{2s_{12}} \,, \\
w^\mu_1 & = p^\mu_1 \,, & 
w^\mu_2 & = p^\mu_2 \,, & 
w^\mu_3 & = \frac{ \spAB{1}{\gamma^\mu}{2} \Phi
          +  \spAB{2}{\gamma^\mu}{1} \Phi^{-1}}{2} \,, &
w^\mu_4 & = \frac{ \spAB{1}{\gamma^\mu}{2} \Phi
          - \spAB{2}{\gamma^\mu}{1} \Phi^{-1}}{2} \,,
\end{align}
where $\Phi$ is an arbitrary phase factor which gives the two terms of $w_3^{\mu}$ and $w_4^{\mu}$ the same helicity scaling. In practice, $\Phi$ cancels out in the hard functions and we can thus set it to $1$.
The five-, four- and three-particle form factors are then obtained from the \emph{contracted} sub-amplitudes
\begin{equation}
	\tilde{A}^{(L)}_{5,uu/ud;i} = u_{i\mu} \, A^{(L)\mu}_{5,uu/ud;i} \,, \qquad
	\tilde{A}^{(L)}_{4,u;i}     = v_{i\mu} \, A^{(L)\mu}_{4,u;i} \,, \qquad
	\tilde{A}^{(L)}_{3;i}       = w_{i\mu} \, A^{(L)\mu}_{3;i} \,, 
\label{eq:contractedamps}
\end{equation}
and the Gram matrices
\begin{equation}
	(\Delta_{5})_{ij} = u_i \cdot u_j \,,\qquad 
	(\Delta_{4})_{ij} = v_i \cdot v_j \,,\qquad 
	(\Delta_{3})_{ij} = w_i \cdot w_j \,,
	\label{eq:GramMatrices}
\end{equation}
as
\begin{align}
	\Omega^{(L)}_{uu/ud;i} & = \sum_{j} \left(\Delta_5^{-1}\right)_{ij} \tilde{A}^{(L)}_{5,uu/ud;j} \,, \\
	\Omega^{(L)}_{u;i}     & = \sum_{j} \left(\Delta_4^{-1}\right)_{ij} \tilde{A}^{(L)}_{4,u;j} \,, \\
	\Omega^{(L)}_{3;i}     & = \sum_{j} \left(\Delta_3^{-1}\right)_{ij} \tilde{A}^{(L)}_{3;j} \,.
\end{align}

\renewcommand{\arraystretch}{1.5}
\begin{table}
\begin{center}
\begin{tabular}{|c|c|	}
\hline 
\# of external particles & independent contracted helicity amplitudes \\
\hline
5 & $\tilde{A}^{(L)+-++}_{5,uu;i} \,, \tilde{A}^{(L)+-+-}_{5,uu;i} \,, \tilde{A}^{(L)+--+}_{5,uu;i} \,, \tilde{A}^{(L)+---}_{5,uu;i}\,,$ \\
  & $\tilde{A}^{(L)+-++}_{5,ud;i} \,, \tilde{A}^{(L)+-+-}_{5,ud;i} \,,$  \\
  & $\tilde{A}^{(L)+-++}_{5,q;i} \,, \tilde{A}^{(L)+-+-}_{5,q;i}$ \\
\hline
4 & $\tilde{A}^{(L)+-+}_{4,u;i}$ \\
\hline
3 & $\tilde{A}^{(L)+-}_{3;i}$ \\
\hline
\end{tabular}
\end{center}
\caption{\label{tab:indepAtilde} Independent helicity configurations for the five-, four- and three-particle amplitudes.}
\end{table}

\begin{figure}[t]
  \begin{center}
    \includegraphics[width=0.65\textwidth]{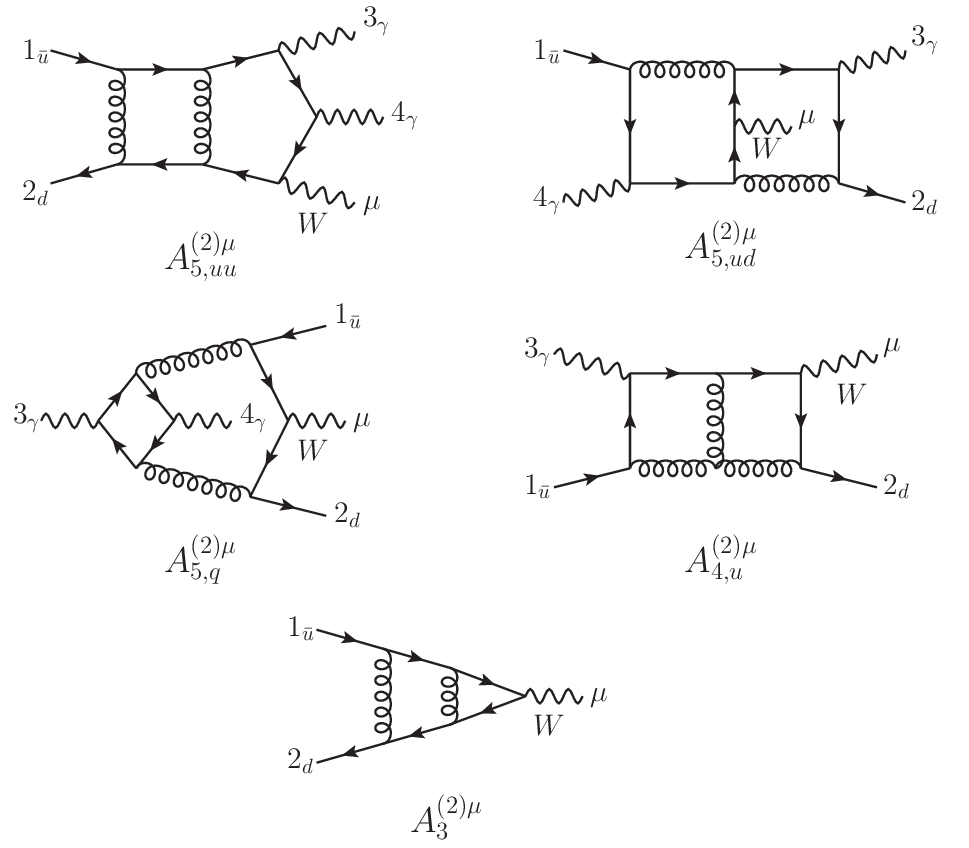}
  \end{center}
	\caption{Sample two-loop Feynman diagrams contributing to the $\wpaa$ independent sub-amplitude currents
		$A^{(2)\mu}_{5,uu}$, $A^{(2)\mu}_{5,ud}$, $A^{(2)\mu}_{5,q}$, $A^{(2)\mu}_{4,u}$ and $A^{(2)\mu}_{3}$.}
  \label{fig:diag2L}
\end{figure}

We then compute the \textit{contracted helicity amplitude}, obtained by specifying the helicity state of the external quarks and photons 
(if they appear in the $W$ production amplitude) in the contracted sub-amplitude $\tilde{A}^{(L)}$. 
In Table~\ref{tab:indepAtilde} we present the independent helicity configurations for five-, four- and three-particle contracted sub-amplitudes.
Moreover, we show sample two-loop Feynman diagrams for the independent five-, four- and three-particle contracted helicity amplitudes in Figure~\ref{fig:diag2L}.

\subsection{Analytic computation framework}
\label{sec:analyticframework}

We now describe our framework to compute the helicity amplitudes for $\waa$ production,
combining Feynman diagrams, IBP reduction, and numerical evaluation over finite fields.
The computation of the contracted helicity amplitudes 
$\tilde{A}^{(L)}_{5,uu/ud;i}$,
$\tilde{A}^{(L)}_{4,u;i}$ and
$\tilde{A}^{(L)}_{3;i}$ defined in Eq.~\eqref{eq:contractedamps} starts with the generation of Feynman diagrams, for which we use \textsc{Qgraf}~\cite{Nogueira:1991ex}.
After identifying the set of distinct denominator structures of the Feynman integrals appearing in the amplitude, 
the numerator of the loop amplitude is written in terms of scalar products of loop and external momenta 
($k_i\cdot k_j$, $k_i \cdot p_j$, $p_i \cdot p_j$). Having the loop-momentum dependent numerator written in this form allows us 
to express the loop amplitude as a linear combination of scalar Feynman integrals, that will further be reduced to a set of master
integral. To achieve such a representation, we follow a strategy that has been used in several previous two-loop five- and four-point amplitude
computations~\cite{Hartanto:2019uvl,Badger:2021owl,Badger:2021imn,Badger:2021ega,Badger:2023mgf,Badger:2023xtl}, 
where we first construct the helicity-dependent loop numerator. This object is obtained by first specifying the helicity 
states of the external quarks and photons, which, in the spinor-helicity formalism, results in the appearance of 
spinor strings which contain loop momenta (e.g.\ $\spAB{i}{\bar{k}_1}{j}$ and $\spAA{i}{\bar{k}_1 \bar{k}_2}{j}$), where $\bar{k}_i$ is the four dimensional component of the loop momentum $k_i$. 
Such objects are cast into a form where the loop-momentum dependence only enters as a scalar product
by expanding the loop momenta using a four-dimensional spanning basis $\eta_j^{\mu}$, $\bar{k}_i^\mu = \sum_{j=1}^{4} a_{ij} \eta_j^\mu$, 
and solving for the coefficients $a_{ij}$, which are functions of $k_i\cdot k_j$, $k_i \cdot p_j$ and $p_i\cdot p_j$.
In addition, the extra dimensional component of the loop momentum ($\tilde{k}_i$) can appear only as $\tilde{k}_i\cdot\tilde{k}_j$, 
which can also be written in terms of $k_i\cdot k_j$, $k_i \cdot p_j$ and $p_i\cdot p_j$.
We express all scalar products and spinor-helicity contractions of the external momenta as rational functions of the five-particle momentum-twistor variables $\vec{x} = (x_1,\ldots,x_6)$ of Ref.~\cite{Badger:2021ega}. 
We repeat their definition in Appendix~\ref{app:MomentumTwistors} for the convenience of the readers.
The processing of loop-momentum dependent helicity numerators is done using a collection of \textsc{Mathematica} and \textsc{Form}~\cite{Ruijl:2017dtg} 
scripts, also with the help of \textsc{Spinney} package~\cite{Cullen:2010jv}.

\smallskip

The computation then proceeds with the IBP reduction of the amplitude onto the master integrals of Refs.~\cite{Abreu:2020jxa,Abreu:2021smk,Abreu:2023rco}.\footnote{We move the square-root normalisations from the master integrals to their expression in terms of special functions.}
The IBP reduction is performed using the strategy outlined in Refs.~\cite{Badger:2021imn,Badger:2023mgf,Badger:2023xtl}, 
which is optimised to handle the large number of permutations 
of the independent integral families that appear in the partial amplitudes. The important element of this technique is that the system of 
IBP equations for each independent family is permuted numerically, which dramatically reduces the memory consumption compared with loading 
a system that incorporates all families ---~including all permutations~--- together. A consequence of this is that master integrals are 
derived for each permutation independently and therefore mappings between sub-topologies must be applied at a second stage before a global 
set of master integrals is obtained. These mappings can be derived using IBP reduction including symmetries, such as those provided by 
\textsc{LiteRed}~\cite{Lee:2013mka}. Even if the global mappings are not applied, the relations will be included once the master integrals have been expanded 
into pentagon functions. The complete computation is stored in a \textsc{FiniteFlow} graph such that all intermediate stages of the reduction 
are numerical.

We generate optimised, small-sized systems of IBP relations using the \textsc{NeatIBP} package~\cite{Wu:2023upw}, which uses the syzygy 
method~\cite{Gluza:2010ws,Chen:2015lyz,Larsen:2015ped,Zhang:2016kfo,Bohm:2017qme,Bosma:2018mtf,Boehm:2020zig}.
Table~\ref{tab:NeatIBP table} shows the \textsc{NeatIBP} performance
in generating the IBP relations.  When generating the IBP relations
using \textsc{NeatIBP}, we found that permuting the propagators
affects the performance, since the syzygy-generator size depends on
the so-called module order.
For family DPzz, we permuted the propagators by hand and obtained the performance shown in Tab.~\ref{tab:NeatIBP table}. 
With the propagator ordering of Ref.~\cite{Abreu:2023rco}, \textsc{NeatIBP} instead generates $71493$ IBP relations with $74143$ relevant integrals, taking $417.0$~MB disk space. 
The corresponding running time is 19h5m (on a computer with 56 CPU threads and 1.5~TB RAM). 
The permutation of the propagators for DPzz decreased the size of the IBP
system by about $14\%$ in terms of the number of IBP relations.
The number of IBP relations and of integrals in the optimised system is much lower as compared to the traditional Laporta 
method.\footnote{We generate the traditional Laporta IBP system with \textsc{LiteRed} by seeding integrals with the maximum rank of 5 and at most one doubled propagator in all sectors.}
This improves the evaluation of the solution to the IBP relations, resulting in both a speed up in the finite-field sampling of the rational coefficients of the amplitudes 
and in a reduction of its memory footprint (by around 8 times and 3 times, respectively, for the leading colour two-loop
five-particle amplitudes).

\renewcommand{\arraystretch}{1.5}
\begin{table}[t!]
\begin{center}
\begin{tabular}{cccccc}
\hline
family &deg.& \# IBPs &\# integrals&IBP disk size&running time\\
\hline
DPmz&5&26673&27432&71.4~MB&7h5m\\

DPzz&5&61777&63880&375.8~MB&21h54m$^*$\\

HBmzz&5&15428&15916&17.0~MB&5h38m\\

HBzmz&5&10953&11289&13.4~MB&5h45m\\

HBzzz&5&21126&21766&38.0~MB&9h32m\\

PBmzz&5&10224&10329&7.8~MB&2h23m\\

PBzmz&5&11610&11791&6.5~MB&2h50m\\

PBzzz&5&8592&8752&5.8~MB&2h50m\\

HTmzzz&4&3120&3176&1.5~MB&1h12m\\

HTzmzz&4&6594&6650&2.5~MB&1h31m\\

HTzzzz&4&4680&4631&4.0~MB&2h31m\\
\hline
\end{tabular}
\end{center}
	\caption{Information about the \textsc{NeatIBP} runs. The first column is the
          name of the integral families following the definition in
          Ref.~\cite{Abreu:2023rco} (the families HTmzzz, HTzmzz
          and HTzzzz only appear in the amplitude computation
          and their definition is shown in Fig.~\ref{fig:HTs}). The
          second column is the corresponding maximum numerator degrees
          of the integrals to be reduced. The next two columns show the
          number of (denoted by \#) IBP relations generated by
          \textsc{NeatIBP}, and of integrals appearing in them. All families are performed on a computer
          with 20 CPU threads and 128~GB RAM, except for DPzz (hence
          the superscript $^*$). For the latter we used a workstation
          with 56 CPU threads and 1.5~TB RAM.}
	\label{tab:NeatIBP table}
\end{table}

\begin{figure}[t!]
  \begin{center}
    \includegraphics[width=0.65\textwidth]{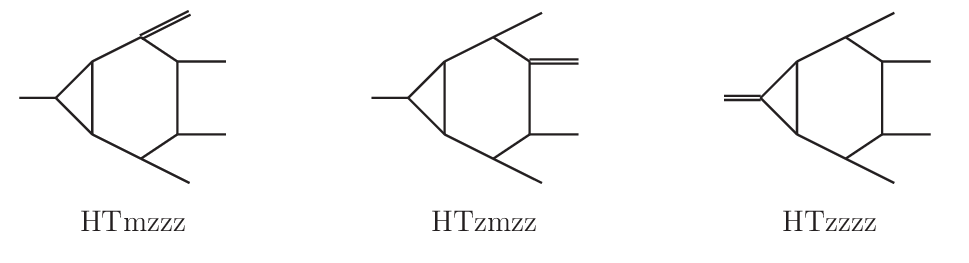}
  \end{center}
	\caption{Pictorial representation of the \textit{hexagon-triangle} integral families entering the two-loop five-particle amplitude computation. The single line represents a massless particle while the double line depicts a massive particle. }
  \label{fig:HTs}
\end{figure}

\smallskip

Once the amplitude is written as a linear combination of master integrals, we perform a Laurent expansion around $\eps = 0$ up to the desired order ($\cO(\eps^0)$ for the two-loop amplitudes, $\cO(\eps^2)$ for the one-loop amplitudes).
In doing this, we express the master integrals in terms of pentagon functions and square roots order by order in $\eps$, using the results of Refs.~\cite{Chicherin:2021dyp,Abreu:2023rco}.
Each bare sub-amplitude $A^{(L)}$ at this stage is expressed as
\begin{equation}
	\label{eq:bareampepsexpanded}
	A^{(L)} = \sum_{s=-2L}^{-2L+4} \; \sum_{r} \; \eps^{s} \; c_{r,s}(\vec{x}) \; m_{r}(f),
\end{equation}
where $m_{r}(f)$ are monomials of pentagon functions $f$, transcendental constants, and square roots, while $c_{r,s}(\vec{x})$ are rational coefficients.
To derive the finite remainders of the contracted helicity amplitudes,
the pole terms in Eqs.~\eqref{eq:pole1L}~and~\eqref{eq:pole2L} are written in terms of pentagon functions and then subtracted from the bare contracted helicity amplitudes.

In our framework, we compute the rational coefficients $c_{r,s}(\vec{x})$ multiplying the pentagon-function monomials in the contracted bare amplitude in Eq.~\eqref{eq:bareampepsexpanded}
by performing numerical evaluations over finite fields within the framework \textsc{FiniteFlow}~\cite{Peraro:2019svx,Peraro:2019okx}. The starting point of this finite-field procedure is the helicity-dependent loop numerator.
The subsequent chain of operations are then carried out numerically over finite fields until we arrive at the $\eps$-expanded contracted bare helicity amplitude representation in Eq.~\eqref{eq:bareampepsexpanded}.
For given input values $\vec{x}$ and prime ${\rm P}$, our numerical algorithm returns the values of the rational coefficients modulus ${\rm P}$.
Their analytic expression is then obtained from several numerical evaluations over finite fields through multivariate functional reconstruction~\cite{Peraro:2016wsq}. 
This approach allows us to obtain analytical results sidestepping the extremely large expressions that would appear in the intermediate stages of the calculation if done symbolically.
We note that, although we could have derived the finite remainders directly, we observe that within our framework the decrease in polynomial degrees going from bare amplitude to finite remainder representations is small
and does not affect the total reconstruction time significantly. 
On that account, we perform analytic reconstruction of the bare helicity amplitudes and afterwards derive the finite remainders.

The reconstruction of the rational coefficients with the generic algorithm implemented in \textsc{FiniteFlow} is however made impractical by the large number of required sample points.
We adopt a strategy used already in a number of similar computations~\cite{Badger:2021nhg,Abreu:2021asb,Badger:2021imn,Badger:2021ega,Badger:2022ncb,Hartanto:2022qhh,Badger:2023mgf,Badger:2023xtl} 
to reduce the number of required sample points.
We refer to Refs.~\cite{Badger:2021imn,Zoia:2023nup} for a more detailed discussion, and give here just a summary.
First of all, we set $x_1 = 1$, and reconstruct only in $x_2, \ldots, x_6$. 
The dependence on $x_1$ is recovered after the reconstruction from dimensional analysis.
The rational coefficients $c_{r,s}(\vec{x})$ in Eq.~\eqref{eq:bareampepsexpanded} with $x_1=1$ are said to be in the `\textbf{original}' stage.
The optimisation of the reconstruction then proceeds in the following stages.
\begin{itemize}
\item[{\bf stage 1}] We fit the $\mathbb{Q}$-linear relations among the $c_{r,s}(\vec{x})$, and solve them so as to choose as independent coefficients the ones with the lowest polynomial degrees.
Only this subset of linearly independent coefficients then needs to be reconstructed.
\item[{\bf stage 2}] We make an ansatz for the denominators of the coefficients as products of factors drawn from the symbol alphabet identified in Refs.~\cite{Abreu:2020jxa,Abreu:2021smk,Abreu:2023rco} and simple spinor brackets.
The exponents of the factors in the ans\"atze are fixed by matching the latter to the coefficients reconstructed on a univariate phase-space slice modulus a prime.
As a result, the denominators are completely determined, and only the numerators remain to be reconstructed.
\item[{\bf stage 3}] We use the information gathered in stage 2 to construct an ansatz in the form of a univariate partial fraction decomposition with respect to $x_5$ for each coefficient, and reconstruct the coefficients in the ansatz. 
By construction, the latter do not depend on $x_5$, so that the reconstruction is now in 4 variables only.\footnote{Evaluating the coefficients of the partial fraction decomposition at a point $(x_2,x_3,x_4,x_6)$ requires evaluating the original coefficients at as many values of $x_5$ as terms in the ansatz (see Refs.~\cite{Peraro:2019svx,Badger:2021imn}). Their evaluation is therefore slower than that of the coefficients of the pentagon-function monomials in Eq.~\eqref{eq:bareampepsexpanded}. Their reconstruction however requires far fewer points so that this strategy is overall advantageous.}
\item[{\bf stage 4}] As in stage~2, we determine the denominators of the coefficients of the partial fraction decomposition through matching on a univariate slice. The partial fractioning introduces spurious denominator factors. These are not included in the original ansatz used at stage~2, but can be derived from it as discussed in Ref.~\cite{Badger:2021imn}.
\end{itemize}
Finally, we use \textsc{FiniteFlow}'s built-in functional reconstruction algorithm to reconstruct the analytic expression of the numerators of the coefficients in the partial fraction decomposition of the coefficients $c_{r,s}(\vec{x})$ in Eq.~\eqref{eq:bareampepsexpanded}.
The numerator/denominator degrees of the coefficients at all stages of the reconstruction strategy for some of the most complicated five-particle partial sub-amplitudes are shown in Table~\ref{tab:reconstructiondata}.

From Table~\ref{tab:reconstructiondata} we can observe that the degree of the rational coefficients increases significantly when we consider the subleading colour amplitudes in comparison to the leading colour ones,
which leads to an increase in the number of sample points required to fully reconstruct the analytic expressions. In addition, the evaluation time and memory consumption also become larger.
We notice that the evaluation time of the subleading colour amplitudes at the stage 4 our reconstruction strategy described above
is about 10 times slower than the leading colour ones, due to the more complicated integrands to be sampled as well as
the larger number of integral family permutations that have to be taken into account in the IBP reduction step. The memory usage was found to be at least 3 times larger, again, due to the substantial size of 
the subleading colour integrands while the non-planar IBP reduction setup does not significantly affect the memory usage since we implement the IBP relations for only one permutation of each integral family.

\smallskip

The univariate partial fractioning is useful to reduce the complexity of the
coefficients but does not necessarily lead to the most compact representations
and in general introduces spurious poles. To further optimise the evaluation
of the analytic expressions we have applied multivariate partial fraction
techniques~\cite{leinartas1978factorization,Boehm:2020ijp,Bendle:2021ueg,Heller:2021qkz}.
The application of these algorithms is particularly challenging due the large number of variables and high polynomial degrees.
In our case, this is further aggravated by the choice of parametrising the kinematics in terms of momentum twistors. 
While this parametrisation has the minimal number of variables and rationalises all spinor
products, the irreducible denominators may have a higher degree compared to
representations with Lorentz invariants.
Sometimes, this makes the computational algebraic geometry computation
in Leinartas-type algorithms and
the `MultivariateApart' algorithm very difficult. 

In order to by-pass these issues, we apply the multivariate partial fractioning to each term in
the univariate partial fractioned representation used in the analytic
reconstruction. Since each term has fewer denominators than the complete set, the
Gr\"obner-basis computation speeds up significantly. 
While this does not remove the spurious factors introduced by the univariate partial fraction decomposition,
it still serves the purpose of making the expressions in general more compact.
We note however that, for a subset of the coefficients, the univariate
decomposition is actually more compact than the multivariate one; in
such cases, we kept the former.
We performed the multivariate partial fraction decompositions by means
of the {\sc pfd-parallel} package~\cite{Bendle:2021ueg}. The
computation was carried out with a workstation using $30$ cores with
$2$~TB RAM for $29.6$ hours. The peak RAM usage for this partial
fraction computation is $17.4$ GB. To further optimise the expressions, we write the separate addends of the partial-fractioned coefficients in terms of independent irreducible factors, which we collect across all amplitudes at each loop order.
The multivariate partial fractioning strategy discussed above leads to
a reduction 
of about $50\%$ of the disk size for the rational coefficients.\footnote{The common factors are collected both before and after the multivariate partial fraction decomposition, to ensure a fair comparison.} 
Moreover, we observe that the evaluation time in \textsc{Mathematica} of all the rational coefficients in the complete finite
remainder decreases by about~20\%.

\renewcommand{\arraystretch}{1.5}
\begin{table}
\begin{center}
\begin{tabular}{cccc|ccccccc}
\hline
		 & original & stage 1 & stage 2 & stage 3  & stage 4 & \makecell{\# points \\ (\# primes)}  & \makecell{analytic \\ expression} \\
\hline
	$\tilde{A}_{5,uu;1}^{(2),N_c^2}$     & 159/155  & 159/155 & 159/0  & 33/31   & 33/0    & 27728 (2) & \cmark \\
	$\tilde{A}_{5,uu;2}^{(2),N_c^2}$     & 147/143  & 147/143 & 147/0  & 33/31   & 33/0    & 37132 (2) & \cmark \\
	$\tilde{A}_{5,uu;3}^{(2),N_c^2}$     & 157/153  & 157/153 & 157/0  & 31/29   & 31/0    & 31610 (2) & \cmark \\
	$\tilde{A}_{5,uu;4}^{(2),N_c^2}$     & 143/139  & 143/139 & 143/0  & 35/33   & 35/0    & 38710 (2) & \cmark \\
\hline
	$\tilde{A}_{5,uu;1}^{(2),1/N_c^2}$     & 223/219  & 223/219 & 223/0  & 50/48   & 50/0    & 134551 (?) & \xmark \\
	$\tilde{A}_{5,uu;2}^{(2),1/N_c^2}$     & 208/204  & 208/204 & 208/0  & 41/42   & 41/0    & 81973 (?)  & \xmark \\
	$\tilde{A}_{5,uu;3}^{(2),1/N_c^2}$     & 219/215  & 219/215 & 219/0  & 49/46   & 49/0    & 130146 (?) & \xmark \\
	$\tilde{A}_{5,uu;4}^{(2),1/N_c^2}$     & 202/199  & 202/199 & 202/0  & 48/49   & 48/0    & 143320 (?) & \xmark \\
\hline
	$\tilde{A}_{5,ud;1}^{(2),N_c^2}$     & 163/160  & 163/160 & 163/0  & 33/32   & 33/0    & 30371 (2) &  \cmark  \\
	$\tilde{A}_{5,ud;2}^{(2),N_c^2}$     & 167/165  & 167/165 & 167/0  & 35/34   & 34/0    & 37506 (2) &  \cmark  \\
	$\tilde{A}_{5,ud;3}^{(2),N_c^2}$     & 150/147  & 150/147 & 150/0  & 33/29   & 31/0    & 29698 (2) &  \cmark \\
	$\tilde{A}_{5,ud;4}^{(2),N_c^2}$     & 152/150  & 152/150 & 152/0  & 35/32   & 34/0    & 36726 (2) &  \cmark \\
\hline
	$\tilde{A}_{5,ud;1}^{(2),1/N_c^2}$     & 219/217  & 217/215 & 217/0  & 55/53   & 55/0    & 173066 (?) &  \xmark \\
	$\tilde{A}_{5,ud;2}^{(2),1/N_c^2}$     & 228/225  & 228/225 & 228/0  & 51/49   & 51/0    & 172337 (?) &  \xmark \\
	$\tilde{A}_{5,ud;3}^{(2),1/N_c^2}$     & 218/213  & 216/211 & 216/0  & 47/45   & 47/0    & 118142 (?) &  \xmark \\
	$\tilde{A}_{5,ud;4}^{(2),1/N_c^2}$     & 208/205  & 206/203 & 206/0  & 50/51   & 50/0    & 153605 (?) &  \xmark \\
\hline
	$\tilde{A}_{5,q;1}^{(2),N_c}$     & 136/135  & 119/118 & 119/0  & 34/32   & 34/0    &  26059 (2) &  \cmark \\
	$\tilde{A}_{5,q;2}^{(2),N_c}$     & 137/137  & 122/122 & 122/0  & 51/52   & 51/0    & 194872 (2) &  \cmark \\
	$\tilde{A}_{5,q;3}^{(2),N_c}$     & 148/147  & 130/129 & 130/0  & 43/44   & 43/0    & 108803 (2) &  \cmark \\
	$\tilde{A}_{5,q;4}^{(2),N_c}$     & 136/135  & 130/129 & 130/0  & 47/48   & 47/0    & 147167 (3) &  \cmark \\
\hline
\end{tabular}
\end{center}
	\caption{\label{tab:reconstructiondata}
	Data about the functional reconstruction.
	The first column lists some of the most complicated independent two-loop five-particle contracted helicity sub-amplitudes.
	From the second to the sixth column, we show the maximal numerator/denominator polynomial degree of the rational coefficients at the various stages of the optimisation discussed in the text.
	For each contracted helicity sub-amplitude, we show the highest degree out of the independent helicity configurations
    listed in Table~\ref{tab:indepAtilde}.
	Before the vertical line, the degrees refer to the coefficients of the pentagon-function monomials ($c_{r,s}(\vec{x})$ in Eq.~\eqref{eq:bareampepsexpanded}) with $x_1=1$; they are functions of 5 variables.
	After the vertical line, they refer to the coefficients in the partial fraction decomposition with respect to $x_5$, which depend on 4 variables.
	The seventh column displays the number of sample points and of prime fields required to complete the functional reconstruction after stage~4 (available only for the amplitudes that were analytically reconstructed). 
	The last column indicates whether the analytic form of the corresponding contracted helicity sub-amplitude is available (\cmark) or not (\xmark).}
\end{table}

\subsection{Numerical approach for the subleading colour five-particle amplitude}
\label{sec:numericalframework}

We have derived analytic expressions for the contracted finite remainders contributing to $\waa$ production at the LHC 
up to two-loop level, except for the subleading colour five-particle $W$-production sub-amplitudes. 
As we have discussed in the previous subsection, the number of sample points required to reconstruct the latter
are considerably higher than for the leading colour ones and the evaluation of the samples is also more expensive in terms of time per point and of memory usage.
In order to complete such analytic reconstructions in a reasonable amount of time one would either need better reconstruction algorithms, or to commit a significant amount of computing resources. 
Therefore, although the latter would be possible in principle,
we believe that committing our computing resources for this computation is not justified.
Several NNLO QCD computations for $2\to 3$ processes have demonstrated that the contribution of the two-loop finite remainders are rather small,
in the range of $\approx 2$~--~$10\%$~\cite{Chawdhry:2019bji,Chawdhry:2021hkp,Kallweit:2020gcp,Czakon:2021mjy,Hartanto:2022qhh,Hartanto:2022ypo,Catani:2022mfv,Buonocore:2022pqq,Buonocore:2023ljm,Mazzitelli:2024ura,Badger:2023mgf}, 
making the contribution of the subleading colour two-loop amplitudes almost negligible. 
In such cases, the leading colour approximation of the two-loop finite remainders is sufficient for a cross section calculation at NNLO QCD. 
Of course, the size of the two-loop finite remainders contributing to the NNLO cross section varies from one process to another, and care must be
taken in estimating the contribution from the subleading colour terms. In addition, the impact of the two-loop subleading colour amplitudes has to be thoroughly assessed also for the 
differential distributions.

We opt to compute numerically the rational coefficients of the pentagon functions appearing in the five-particle $W$-production two-loop finite remainders.
The subleading colour four- and three- particle sub-amplitudes, however, are computed analytically, since they are comparatively simpler.
While obtaining results using this numerical approach is considerably more expensive as compared to evaluating analytic expressions, 
we expect that it will still be feasible to include the subleading colour contributions of the two-loop finite remainders by means of re-weighting~\cite{Bevilacqua:2011xh,Chawdhry:2019bji}. 

For this numerical evaluation, we employ the analytic framework described in Section~\ref{sec:analyticframework}, albeit with a number of modifications.
First of all, in our analytic computation of the helicity amplitudes we parametrise the external kinematics in terms momentum-twistor variables. 
For a physical phase-space point, some of the momentum twistor variables ($x_2,x_3,x_5$) are complex valued, this way making our setup incompatible with the finite-field approach as is. 
To overcome this issue, we employ the four-dimensional projector method~\cite{Peraro:2019cjj,Peraro:2020sfm}
to construct the unreduced amplitude. In this construction, we can work directly with the Mandelstam invariants, which are by definition real valued.
In order to build the unreduced amplitude, we first consider the tensor decomposition of the contracted amplitudes in Eq.~\eqref{eq:contractedamps},
\begin{equation}
	\tilde{A}_{5,uu/ud,i}^{(L)} = \sum_{j=1}^{8} \chi_{uu/ud,i;j}^{(L)} \, T_j  \,,
	\label{eq:tensordecomposition1}
\end{equation}
with
\begin{align}
\label{eq:tensordecomposition2} 
\begin{aligned}
	T_1 & = \bar{u}(p_1) \slashed{p}_3 v(p_2) \; p_1 \cdot \varepsilon(p_3,q_3) \; p_1 \cdot \varepsilon(p_4,q_4) \,,  \\ 
	T_2 & = \bar{u}(p_1) \slashed{p}_3 v(p_2) \; p_1 \cdot \varepsilon(p_3,q_3) \; p_2 \cdot \varepsilon(p_4,q_4) \,,  \\ 
	T_3 & = \bar{u}(p_1) \slashed{p}_3 v(p_2) \; p_2 \cdot \varepsilon(p_3,q_3) \; p_1 \cdot \varepsilon(p_4,q_4) \,,  \\ 
	T_4 & = \bar{u}(p_1) \slashed{p}_3 v(p_2) \; p_2 \cdot \varepsilon(p_3,q_3) \; p_2 \cdot \varepsilon(p_4,q_4) \,,  \\ 
	T_5 & = \bar{u}(p_1) \slashed{p}_4 v(p_2) \; p_1 \cdot \varepsilon(p_3,q_3) \; p_1 \cdot \varepsilon(p_4,q_4) \,,  \\ 
	T_6 & = \bar{u}(p_1) \slashed{p}_4 v(p_2) \; p_1 \cdot \varepsilon(p_3,q_3) \; p_2 \cdot \varepsilon(p_4,q_4) \,,  \\ 
	T_7 & = \bar{u}(p_1) \slashed{p}_4 v(p_2) \; p_2 \cdot \varepsilon(p_3,q_3) \; p_1 \cdot \varepsilon(p_4,q_4) \,,  \\ 
	T_8 & = \bar{u}(p_1) \slashed{p}_4 v(p_2) \; p_2 \cdot \varepsilon(p_3,q_3) \; p_2 \cdot \varepsilon(p_4,q_4) \,,
\end{aligned}
\end{align}
where $q_i$ are arbitrary reference momenta.
We choose them as $q_3=p_4$ and $q_4=p_3$.
We stress that the same choice of photon's reference vectors will be used throughout the whole calculation, i.e.\ in the computation of the five- and four-particle sub-amplitudes
as well as of the decay currents.
The form factors are then obtained by
\begin{equation}
	\chi_{uu/ud,k;i}^{(L)} = \sum_{j} \left( \Theta^{-1} \right)_{ij} \; T^{\dagger}_{j} \cdot \tilde{A}_{5,uu/ud,k}^{(L)} \,,
\label{eq:contractedampsij}
\end{equation}
where $\Theta_{ij} = T^{\dagger}_{i} \cdot T_{j}$. 
We derive the analytic form of the unreduced $T^{\dagger}_{j} \cdot \tilde{A}_{5,uu/ud,k}^{(L)}$ amplitude as a linear combination of 
scalar Feynman integrals. 
This is the starting point of our finite-field computation. The subsequent steps ---~IBP reduction, $\eps$-expansion and map to pentagon functions~--- 
are carried out numerically over finite field as discussed in Section~\ref{sec:analyticframework}. 
Instead of performing a univariate partial fraction decomposition, however,
we substitute the rationalised numerical values of the Mandelstam invariants. 
For each phase-space point we need to perform the numerical evaluation on several finite fields ---~that is, modulo several different primes~--- in order to reconstruct 
the numerical values of the rational coefficients. A similar approach, where rational kinematics is employed in the IBP reduction, has been used for example in Refs.~\cite{Bronnum-Hansen:2021olh,Agarwal:2024jyq}.

We present in Table~\ref{tab:degreedatasij} the polynomial degree information on the rational coefficients of the pentagon functions appearing in $T^{\dagger}_{j} \cdot \tilde{A}_{5,uu/ud,k}^{(L)}$ at stage~2 (after imposing linear relations and denominator guessing) prior to the numerical reconstruction.
Finally, we derive the contracted helicity amplitudes ($\tilde{A}_{5,uu/ud,i}^{(L)}$) numerically by specifying the helicity states of the external particles
in Eq.~\eqref{eq:tensordecomposition2} and plugging them into Eq.~\eqref{eq:tensordecomposition1}, to be combined with the form factors $\chi_{uu/ud,i;j}^{(L)}$ numerically.

\renewcommand{\arraystretch}{1.5}
\begin{table}
\begin{center}
\begin{tabular}{cccc}
\hline
		 & original & stage 1 & stage 2  \\
\hline
	$T^{\dagger}_{j} \cdot \tilde{A}_{5,uu,1}^{(2),1/N_c^2}$     & 100/97  & 100/97 & 100/0   \\
	$T^{\dagger}_{j} \cdot \tilde{A}_{5,uu,2}^{(2),1/N_c^2}$     &  99/96  &  99/96 &  99/0   \\
	$T^{\dagger}_{j} \cdot \tilde{A}_{5,uu,3}^{(2),1/N_c^2}$     & 101/97  & 101/97 & 101/0   \\
	$T^{\dagger}_{j} \cdot \tilde{A}_{5,uu,4}^{(2),1/N_c^2}$     & 101/97  & 101/97 & 101/0   \\
\hline
	$T^{\dagger}_{j} \cdot \tilde{A}_{5,ud,1}^{(2),1/N_c^2}$     &  97/93  &  96/92 &  96/0   \\
	$T^{\dagger}_{j} \cdot \tilde{A}_{5,ud,2}^{(2),1/N_c^2}$     &  97/93  &  97/93 &  97/0   \\
	$T^{\dagger}_{j} \cdot \tilde{A}_{5,ud,3}^{(2),1/N_c^2}$     &  97/93  &  96/92 &  96/0   \\
	$T^{\dagger}_{j} \cdot \tilde{A}_{5,ud,4}^{(2),1/N_c^2}$     &  97/93  &  96/92 &  96/0   \\
\hline
\end{tabular}
\end{center}
	\caption{\label{tab:degreedatasij} Maximal numerator/denominator polynomial degree of the rational coefficients of the pentagon-function monomials appearing in the $T^{\dagger}_{j} \cdot \tilde{A}_{5,uu/ud,i}^{(2),1/N_c^2}$ amplitude up to stage~2
    of our optimisation strategy. We note that $T^{\dagger}_{j} \cdot \tilde{A}_{5,uu/ud,i}^{(2),1/N_c^2}$ are functions of the five-particle
    Mandelstam invariants~$\vec{s}_5$.}
\end{table}

\subsection{Checks}

We have performed the following checks to validate our results.
\begin{itemize}
\item \textit{Gauge invariance} \\
We verified the gauge invariance by checking that the Ward identities are satisfied, i.e., that replacing one of the photon polarisation vectors by its momentum yields zero.
We recall that the individual terms in the amplitude decomposition in Eq.~\ref{eq:ampdecomposition} are not separately
gauge invariant, and checking the Ward identities thus requires summing all the contributions.
Furthermore, we use the complex mass scheme to ensure that the Ward identities are satisfied~\cite{Baur:1997bn,Denner:1999gp,Denner:2005fg,Denner:2006ic}. 
We evaluated the rational coefficients of the full colour amplitude numerically at a non-physical rational phase-space point (where all the momentum-twistor variables are real and positive) and left the pentagon functions symbolic. 
This is possible because the pentagon functions are by construction algebraically independent.

\item \textit{Cancellation of UV and IR singularities} \\
In deriving the finite remainders, we demonstrate that the UV and IR poles cancel out as expected. 
This serves as a strong consistency check of our loop-amplitude computation. While we demonstrated the pole cancellation analytically
for the leading colour amplitudes, for the subleading colour contributions we checked the cancellation of the poles numerically at a non-physical 
rational phase-space point (again leaving the pentagon functions symbolic).

\item \textit{Renormalisation-scale dependence} \\
We checked that the numerical evaluation of the finite remainders at $\mu_R \ne 1$, obtained through Eqs.~\eqref{eq:mudepNcp2}~--~\eqref{eq:mudepNcNf}, agrees with the evaluation at $\mu_R=1$ and rescaled values of the momenta as dictated by dimensional analysis,
\begin{equation}
\frac{F^{(L)}_{6}(\vec{p},\mu_R=a)}{A^{(0)}_{6}(\vec{p})} = \frac{F^{(L)}_{6}(\vec{p}/a,\mu_R=1)}{A^{(0)}_{6}(\vec{p}/a)} \,,
\end{equation}
where $\vec{p}$ is a shorthand for the list of all external momenta ($p_1,p_2,p_3,p_4,p_5,p_6$). 
We note that the tree-level amplitude does not depend on $\mu_R$. 

\item \textit{One-loop comparison against \textsc{OpenLoops}} \\
We compared against \textsc{OpenLoops}~\cite{Buccioni:2019sur} at the level of the squared bare amplitudes at tree and one-loop levels, in full colour, for both the $\wpaa$ and $\wmaa$ processes. 
While we derived the one-loop amplitude through $\cO(\eps^2)$, the comparison against \textsc{OpenLoops} is carried out only up to $\cO(\eps^0)$.

\end{itemize}

\section{Results}
\label{sec:result}

We derived analytic expressions for the leading colour two-loop amplitudes required to obtain NNLO QCD predictions for $pp \to W\gamma\gamma$ production at the LHC.
For the subleading colour contributions, we set up a routine which allows us to obtain numerical values, albeit with a significant computational cost. 
Both the analytic and numerical results rely on the computational framework based on finite-field arithmetic presented in the previous section. 
In this section, we first present benchmark numerical results for the bare helicity amplitude and hard functions 
at a phase-space point in the physical scattering region. We further discuss the analytic properties of the two-loop amplitudes and describe the ancillary files that accompany this paper.

\subsection{Numerical results}

At parton level, $W^{+}\gamma\gamma$ production at hadron colliders consists of the following two scattering channels,
\begin{align}
\label{eq:wpaachannel}
\begin{aligned}
	&  u(-p_1) + \bar{d}(-p_2) \to \gamma(p_3) + \gamma(p_4) + \nu_{\ell}(p_5) + \ell^+(p_6) \,, \\
	&  \bar{d}(-p_1) + u(-p_2) \to \gamma(p_3) + \gamma(p_4) + \nu_{\ell}(p_5) + \ell^+(p_6) \,, \\
\end{aligned}
\end{align}
while for $W^{-}\gamma\gamma$ production we have
\begin{align}
\label{eq:wmaachannel}
\begin{aligned}
	&  d(-p_1) + \bar{u}(-p_2) \to \gamma(p_3) + \gamma(p_4) + \ell^-(p_5) + \bar{\nu}_{\ell}(p_6) \,, \\
	&  \bar{u}(-p_1) + d(-p_2) \to \gamma(p_3) + \gamma(p_4) + \ell^-(p_5) + \bar{\nu}_{\ell}(p_6) \,. \\
\end{aligned}
\end{align}
We present benchmark values at the following phase-space point in the physical scattering region:
\begin{align}
\label{eq:PSpoint}
\begin{gathered}
\begin{alignedat}{4}
	s_{12} & =  10^6 \; \mathrm{GeV}^2 \,, \qquad
	\frac{s_{23}}{s_{12}} &&  = -\frac{1}{78} \,, \qquad
	\frac{s_{34}}{s_{12}}  && = \frac{5}{56}  \,, \qquad
	\frac{s_{45}}{s_{12}}  &&  = \frac{1}{87}  \,, \\
	\frac{s_{56}}{s_{12}}   & = \frac{23}{90} \,,  \qquad \qquad \quad
	\frac{s_{16}}{s_{12}}   && = -\frac{38}{79} \,, \qquad
	\frac{s_{123}}{s_{12}}  && =  \frac{79}{96} \,, \qquad
	\frac{s_{234}}{s_{12}}  && = -\frac{21}{64} \,, 
\end{alignedat}\\
\tr_5(1234)  = - \i \frac{\sqrt{300731287031}}{25159680}  \; \mathrm{GeV}^2 \,. 
\end{gathered}
\end{align}
We refer to Appendix~\ref{app:rationalise6pt} for the definition of the parity-odd invariant $\tr_5(1234)$.
The other input parameters are
\begin{equation}
\begin{alignedat}{2}
 M_W &= 80.419~\GeV\,, \qquad \Gamma_W &&= 2.0476~\GeV\,, \\
 M_Z &= 91.188~\GeV\,, \qquad \Gamma_Z &&= 2.441404~\GeV\,, \\
\alpha_s &= 0.118 \,, \qquad \qquad \quad \alpha^{-1} &&= 137.035999 \,, \\
	n_f &= 5 \,.
\end{alignedat}
\label{eq:inputparameters}
\end{equation}
We employ the complex mass scheme to preserve the gauge invariance. In this scheme,
the masses of the electroweak vector bosons ($\mu_W$ and $\mu_Z$) and the weak coupling ($g_W$) are complex valued:
\begin{align}
	\mu_W^2 & = M_W^2 - \i \, \Gamma_W \, M_W \,,\\
	\mu_Z^2 & = M_Z^2 - \i \,  \Gamma_Z \, M_Z  \,, \\
	g_W     & = \frac{e}{\sqrt{1-\frac{\mu_W^2}{\mu_Z^2}}} \,.
\end{align}

In order to obtain the full set of six-point helicity amplitudes specified in Eq.~\eqref{eq:sixpthelicityamps} for all contributions appearing in 
Eq.~\eqref{eq:ampdecomposition} from the independent five-, four- and three-particle helicity amplitudes tabulated in Table~\ref{tab:indepAtilde}, 
we need to apply permutations of the external momenta and/or parity transformation.
The permutation of the momentum-twistor variables $\vec{z}$ and $\vec{x}$ is achieved by permuting the external momenta on the RHS of 
their definition in Eqs.~\eqref{eq:momtwistor6pt}~and~\eqref{eq:momtwistor5pt}.
This suffices to obtain the values of the rational coefficients upon momentum permutation.\footnote{This operation must be performed on expressions that are free of helicity phases, which can be obtained by normalising each helicity sub-amplitude by an arbitrary phase factor. See Appendix~C of Ref.~\cite{Badger:2023mgf} for a detailed discussion of this.}
Permuting the momenta in the pentagon functions requires instead some work, as they can only be evaluated in the $s_{12}$ channel.\footnote{\label{note-momenta}Note the different labelling of the momenta. 
Our momenta ($p_i$) are related to the ones used in the pentagon functions ($p_i'$)~\cite{Chicherin:2021dyp,Abreu:2023rco} by $\bigl(p_1', p_2', p_3', p_4', p_5'\bigr) = \bigl(p_5 + p_6, p_4, p_3, p_2, p_1\bigr)$.}
In order to evaluate the pentagon functions in different channels, we rely on the fact that the pentagon functions cover the master integrals in all permutations of the external massless legs.
This in fact allows us to express the pentagon functions evaluated in any physical channel as polynomials in pentagons functions evaluated in the $s_{12}$ channel.
These relations can be obtained by rewriting the pentagon functions in terms of master-integral components, permuting the external massless legs in them, and re-expressing the result in terms of pentagon functions.
We provide them in our ancillary files.
We refer e.g.\ to Refs.~\cite{Chicherin:2021dyp,Badger:2022ncb,Badger:2023mgf} for a more thorough discussion of how to permute the pentagon functions.
We emphasise that, with this approach, evaluating the pentagon functions at a single phase-space point in the $s_{12}$ channel suffices to cover all possible crossings of the integrals/amplitudes.

In Tables~\ref{tab:benchmark1Lbare}~and~\ref{tab:benchmark2Lbare}, we present numerical results for the bare one- and two-loop six-point helicity amplitudes normalised to the tree-level amplitude, 
\begin{equation}
	\hat{A}^{(L),i}_{6} = \frac{A^{(L),i}_{6}}{A^{(0),i}_{6}} \,,
\end{equation}
for the $u\bar{d}\to\gamma\gamma\nu_\ell \ell^+$ scattering channel evaluated at the phase-space point defined in Eq.~\eqref{eq:PSpoint} 
for all $N_c$ and $n_f$ components listed in Eqs.~\eqref{eq:NcNfdecomposition1L}~--~\eqref{eq:NcNfdecomposition2Lqq},
with $\mu_R = 1~\GeV$. Furthermore, we show the numerical results of the hard functions defined in Eqs.~\eqref{eq:hardfunction0L}~--~\eqref{eq:hardfunction2L}
for all scattering channels contributing to $\wpaa$ and $\wmaa$ productions in 
Tables~\ref{tab:num-hardfunction-muR1}~and~\ref{tab:num-hardfunction-muR100} with the renormalisation scale $\mu_R$ 
is set to 1 and 100 GeV, respectively. 

\renewcommand{\arraystretch}{1.5}
\begin{table}[t!]
\centering
\begin{tabularx}{1.0\textwidth}{|C{0.7}|C{1.0}|C{0.6}|C{1.0}|C{1.2}|C{1.2}|C{1.3}|}
\hline
 $\wpaa$     & helicity & $\eps^{-2}$ & $\eps^{-1}$ & $\eps^{0}$ & $\eps^{1}$ & $\eps^{2}$ \\
\hline
$\hat A^{(1),N_c}_{6}$ & $\scriptstyle +-++-+$ & $ -1 $ & $ 12.3155 - 3.14159 \i $ & $ -70.4776 + 36.4764 \i $ & $ 248.909 - 210.537 \i $ & $ -594.200 + 804.874 \i $ \\
                       & $\scriptstyle +-+--+$ & $ -1 $ & $ 12.3155 - 3.14159 \i $ & $ -73.4410 + 37.2798 \i $ & $ 280.391 - 220.225 \i $ & $ -763.959 + 864.563 \i $ \\
                       & $\scriptstyle +--+-+$ & $ -1 $ & $ 12.3155 - 3.14159 \i $ & $ -71.0908 + 35.1241 \i $ & $ 255.807 - 197.895 \i $ & $ -633.056 + 745.050 \i $ \\
                       & $\scriptstyle +----+$ & $ -1 $ & $ 12.3155 - 3.14159 \i $ & $ -73.1768 + 37.0202 \i $ & $ 278.485 - 217.118 \i $ & $ -758.473 + 848.321 \i $ \\
\hline
\end{tabularx}
\caption{\label{tab:benchmark1Lbare} Bare one-loop six-point helicity amplitudes normalised to the tree-level amplitudes for $\wpaa$ production
in the $u\bar{d}\to\gamma\gamma\nu_\ell \ell^+$ scattering channel evaluated at the phase-space point given in Eq.~\eqref{eq:PSpoint} with $\mu_R = 1~\mathrm{GeV}$.
}
\end{table}

\renewcommand{\arraystretch}{1.5}
\begin{table}[t!]
\centering
\begin{tabularx}{1.0\textwidth}{|C{0.7}|C{1.0}|C{0.6}|C{1.0}|C{1.2}|C{1.2}|C{1.3}|}
\hline
 $\wpaa$     & helicity & $\eps^{-4}$ & $\eps^{-3}$ & $\eps^{-2}$ & $\eps^{-1}$ & $\eps^{0}$ \\
\hline
$\hat A^{(2),N_c^2}_{6}$ & $\scriptstyle +-++-+$ & $ 0.5 $ & $ -13.2322 + 3.14159 \i $ & $ 162.918 - 80.9263 \i $ & $ -1234.49 + 1008.40 \i $ & $ 6330.53 - 8136.93 \i $ \\
                         & $\scriptstyle +-+--+$ & $ 0.5 $ & $ -13.2322 + 3.14159 \i $ & $ 165.882 - 81.7297 \i $ & $ -1310.81 + 1040.23 \i $ & $ 7274.97 - 8660.09 \i $ \\
                         & $\scriptstyle +--+-+$ & $ 0.5 $ & $ -13.2322 + 3.14159 \i $ & $ 163.532 - 79.5740 \i $ & $ -1255.43 + 976.066 \i $ & $ 6636.25 - 7775.07 \i $ \\
                         & $\scriptstyle +----+$ & $ 0.5 $ & $ -13.2322 + 3.14159 \i $ & $ 165.618 - 81.4701 \i $ & $ -1305.49 + 1032.15 \i $ & $ 7230.30 - 8550.74 \i $ \\
\hline
$\hat A^{(2),1/N_c^2}_{6}$ & $\scriptstyle +-++-+$ & $ 0.5 $ & $ -12.3155 + 3.14159 \i $ & $ 141.379 - 75.1667 \i $ & $ -1003.61 + 881.175 \i $ & $ 4845.29 - 6798.85 \i $ \\
                           & $\scriptstyle +-+--+$ & $ 0.5 $ & $ -12.3155 + 3.14159 \i $ & $ 144.342 - 75.9701 \i $ & $ -1069.06 + 910.066 \i $ & $ 5551.74 - 7191.47 \i $ \\
                           & $\scriptstyle +--+-+$ & $ 0.5 $ & $ -12.3155 + 3.14159 \i $ & $ 141.992 - 73.8144 \i $ & $ -1022.31 + 853.804 \i $ & $ 5082.41 - 6537.13 \i $ \\
                           & $\scriptstyle +----+$ & $ 0.5 $ & $ -12.3155 + 3.14159 \i $ & $ 144.078 - 75.7105 \i $ & $ -1064.72 + 902.932 \i $ & $ 5526.61 - 7100.16 \i $ \\
\hline
$\hat A^{(2),N_c n_f}_{6}$ & $\scriptstyle +-++-+$ & $ 0 $ & $ 0.166667 $ & $ -3.82739 + 1.04720 \i $ & $ 40.7341 - 22.5723 \i $ & $ -267.947 + 244.573 \i $ \\
                           & $\scriptstyle +-+--+$ & $ 0 $ & $ 0.166667 $ & $ -3.82739 + 1.04720 \i $ & $ 42.7097 - 23.1079 \i $ & $ -307.262 + 256.430 \i $ \\
                           & $\scriptstyle +--+-+$ & $ 0 $ & $ 0.166667 $ & $ -3.82739 + 1.04720 \i $ & $ 41.1429 - 21.6707 \i $ & $ -276.275 + 228.505 \i $ \\
                           & $\scriptstyle +----+$ & $ 0 $ & $ 0.166667 $ & $ -3.82739 + 1.04720 \i $ & $ 42.5336 - 22.9348 \i $ & $ -304.621 + 252.861 \i $ \\
\hline
$\hat A^{(2),N_c}_{6,q}$ & $\scriptstyle +-++-+$ & $ 0 $ & $ 0 $ & $ 0 $ & $ 0 $ & $ -18.4450 - 44.8624 \i $ \\
                         & $\scriptstyle +-+--+$ & $ 0 $ & $ 0 $ & $ 0 $ & $ 0 $ & $ 15.8310 + 23.0822 \i $ \\
                         & $\scriptstyle +--+-+$ & $ 0 $ & $ 0 $ & $ 0 $ & $ 0 $ & $ 16.5070 - 8.58631 \i $ \\
                         & $\scriptstyle +----+$ & $ 0 $ & $ 0 $ & $ 0 $ & $ 0 $ & $ 0.0820702 + 1.03421 \i $ \\
\hline
\end{tabularx}
\caption{\label{tab:benchmark2Lbare} 
Bare two-loop six-point helicity amplitudes normalised to the tree-level amplitudes for $\wpaa$ production
in the $u\bar{d}\to\gamma\gamma\nu_\ell \ell^+$ scattering channel evaluated at the phase-space point given in Eq.~\eqref{eq:PSpoint} with $\mu_R = 1~\mathrm{GeV}$.
We show all $N_c$ and $n_f$ components appearing in Eqs.~\eqref{eq:NcNfdecomposition2L}~and~\eqref{eq:NcNfdecomposition2Lqq}.
Here the $A_{6,q}^{(2)}$ has been separated out from the other sub-amplitudes contributing to the six-point amplitude 
as given in Eq.~\eqref{eq:ampdecomposition}.
}
\end{table}

\renewcommand{\arraystretch}{1.5}
\begin{table}
    \centering
	\begin{tabularx}{1\textwidth}{|C{0.9}|C{1.1}|C{1.0}|C{1.0}|C{1.0}|}
    \hline
	    $\wpaa$ & $\cH^{(0)}$ [GeV$^{-4}$] & $\cH^{(1)}/\cH^{(0)}$ & $\cH^{(2)}/\cH^{(0)}$ & $\cH^{(2)}_{\lc}/\cH^{(0)}$  \\
    \hline
	    $u\bar{d}\to \gamma\gamma \nu_\ell \ell^+$  & $ 5.111840  \cdot 10^{-14}$ & -0.2048115 & 0.09738631 & 0.1171781  \\
	    $\bar{d}u\to \gamma\gamma \nu_\ell \ell^+$  & $ 0.7177927 \cdot 10^{-14}$ & -0.3402089 & 0.2266332  & 0.2639021  \\
    \hline
	    $\wmaa$ & $\cH^{(0)}$ [GeV$^{-4}$]& $\cH^{(1)}/\cH^{(0)}$ & $\cH^{(2)}/\cH^{(0)}$& $\cH^{(2)}_{\lc}/\cH^{(0)}$  \\
    \hline
	    $d\bar{u}\to \gamma\gamma \bar{\nu}_\ell \ell^-$  & $3.906753  \cdot 10^{-12}$ & -0.1711305 & 0.05641518 & 0.07114385 \\
	    $\bar{u}d\to \gamma\gamma \bar{\nu}_\ell \ell^-$  & $0.367502  \cdot 10^{-12}$ & -0.1508495 & 0.03488613 & 0.04697647 \\
    \hline
    \end{tabularx}
	\caption{Benchmark numerical results for the tree-level, one- and two-loop hard functions 
    (defined in Eqs.~\eqref{eq:hardfunction0L}~--~\eqref{eq:hardfunction2L}) for all scattering channels contributing
    to $\wpaa$ and $\wmaa$ productions at the phase-space point given in Eq.~\eqref{eq:PSpoint} with $\mu_R = 1~\mathrm{GeV}$. 
    For the two-loop hard functions we also provide the results in the leading-colour approximation. The one- and two-loop amplitude
    decompositions in the leading colour approximation are given in Eqs.~\eqref{eq:NcNfdecomposition1Llc}~--~\eqref{eq:NcNfdecomposition2Lqqlc}.
    }
    \label{tab:num-hardfunction-muR1}
\end{table}

\renewcommand{\arraystretch}{1.5}
\begin{table}
    \centering
	\begin{tabularx}{1\textwidth}{|C{0.9}|C{1.1}|C{1.0}|C{1.0}|C{1.0}|}
    \hline
	    $\wpaa$ & $\cH^{(0)}$ [GeV$^{-4}$] & $\cH^{(1)}/\cH^{(0)}$ & $\cH^{(2)}/\cH^{(0)}$ & $\cH^{(2)}_{\lc}/\cH^{(0)}$  \\
    \hline
	    $u\bar{d}\to \gamma\gamma \nu_\ell \ell^+$  & $ 5.111840  \cdot 10^{-14}$ & -0.2048115 &  0.04836010 & 0.05771527  \\
	    $\bar{d}u\to \gamma\gamma \nu_\ell \ell^+$  & $ 0.7177927 \cdot 10^{-14}$ & -0.3402089 &  0.08783011 & 0.1034402   \\
    \hline
	    $\wmaa$ & $\cH^{(0)}$ [GeV$^{-4}$]& $\cH^{(1)}/\cH^{(0)}$ & $\cH^{(2)}/\cH^{(0)}$& $\cH^{(2)}_{\lc}/\cH^{(0)}$  \\
    \hline
	    $d\bar{u}\to \gamma\gamma \bar{\nu}_\ell \ell^-$  & $3.906753  \cdot 10^{-12}$ & -0.1711305 & 0.02972155 & 0.03680515 \\
	    $\bar{u}d\to \gamma\gamma \bar{\nu}_\ell \ell^-$  & $0.367502  \cdot 10^{-12}$ & -0.1508495 & 0.02164006 & 0.02776627 \\
    \hline
    \end{tabularx}
	\caption{Same as Table~\ref{tab:num-hardfunction-muR100} but with $\mu_R = 100~\mathrm{GeV}$. 
    }
    \label{tab:num-hardfunction-muR100}
\end{table}

To obtain numerical results at full colour for both bare helicity amplitudes and hard functions, we need to determine the rational coefficients of the pentagon functions for the subleading
colour two-loop five-particle amplitudes numerically using the approach discussed in Section~\ref{sec:numericalframework}. To derive numerical results for both the $\wpaa$ and $\wmaa$ productions
for all scattering channels shown in 
Eqs.~\eqref{eq:wpaachannel}~and~\eqref{eq:wmaachannel}, we need to compute the $T^{\dagger}_{j} \cdot \tilde{A}_{5,uu,i}^{(2),1/N_c^2}$ and $T^{\dagger}_{j} \cdot \tilde{A}_{5,ud,i}^{(2),1/N_c^2}$ amplitudes 
(defined in Eq.~\eqref{eq:contractedampsij} with $i=1,\cdots,32$) evaluated for the five-particle Mandelstam invariants 
$\vec{s}_5$ and $\vec{s}^{\,\prime}_5$ where $\vec{s}^{\,\prime}_5 = \vec{s}_{5}|_{p_1\leftrightarrow p_2}$.
We in fact recall that the momentum swap $p_1 \leftrightarrow p_2$ allows us to obtain $\bar{d}u\to\gamma\gamma\nu_\ell \ell^+$ ($\bar{u}d\to\gamma\gamma\ell^- \bar{\nu}_\ell$) 
from $u\bar{d}\to\gamma\gamma\nu_\ell \ell^+$ ($d\bar{u}\to\gamma\gamma\ell^- \bar{\nu}_\ell$)
as well as the $\wmaa$ amplitudes from the $\wpaa$ ones.
The evaluation time for the $T^{\dagger}_{j} \cdot \tilde{A}_{5,uu,i}^{(2),1/N_c^2}$ amplitudes at the stage 2 of our optimisation strategy (described in Section~\ref{sec:analyticframework}) 
per phase-space point and for a single prime field is about 50 
seconds, while for  $T^{\dagger}_{j} \cdot \tilde{A}_{5,ud,i}^{(2),1/N_c^2}$ amplitudes it is about 85 seconds.\footnote{In our setup, the $T^{\dagger}_{j} \cdot \tilde{A}_{5,uu/ud,i}^{(2),1/N_c^2}$ amplitudes are processed simultaneously for all the tensor structures ($j=1,\cdots,8$)
and for each $i = 1,\cdots,4$.
The evaluation times quoted in this work are measured on an Intel(R) Xeon(R) Gold 6432 2.80 GHz CPU.} 
At least 33 prime fields are needed to fully reconstruct the values of rational coefficient of the pentagon functions at the phase-space point specified in
Eq.~\eqref{eq:PSpoint}.

\begin{figure}[t!]
\begin{subfigure}{.5\textwidth}
\centering
\includegraphics[width=0.9\textwidth]{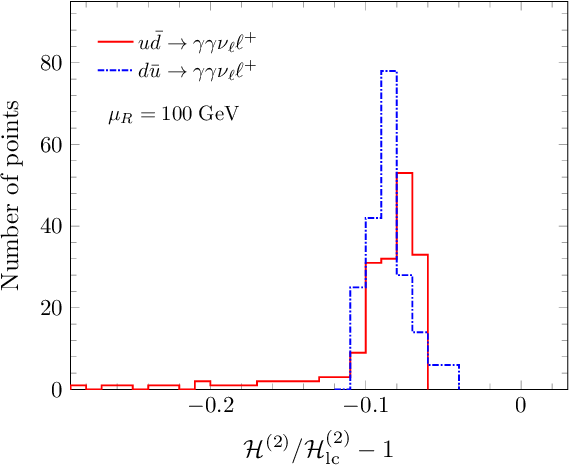}
\end{subfigure}
\begin{subfigure}{.5\textwidth}
\centering
\includegraphics[width=0.9\textwidth]{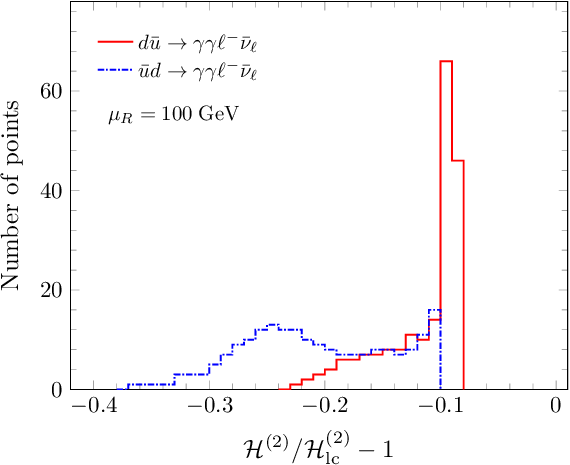}
\end{subfigure}
\caption{Distribution of subleading colour corrections to the two-loop hard functions for the set of phase-space point defined in 
	Eqs.~\eqref{eq:unislice1}~and~\eqref{eq:unisliceparams}, for $\wpaa$ (left) and $\wmaa$ (right) production.}
\label{fig:hardfunctions_slc_corrections}
\end{figure}

To demonstrate the feasibility of our approach to compute the full-colour amplitude for a relatively large set of phase-space points, we evaluated both the 
$\wpaa$ and $\wmaa$ hard functions on a univariate phase-space slice,
parametrised as
\begin{align} 
\label{eq:unislice1}
\begin{aligned}
& p_1^{\mu} =  \frac{\sqrt{s}}{2} \, \left(-1,0,0,-1 \right) \,, \\
& p_2^{\mu} =  \frac{\sqrt{s}}{2} \, \left(-1,0,0,1 \right) \,, \\
& p_3^{\mu} = u_1 \, \frac{\sqrt{s}}{2} \, \left(1,t,\frac{t}{1000},\frac{t}{1000} \right) \,, \\
& p_4^{\mu} = u_2 \, \frac{\sqrt{s}}{2} \, \left(1,\cos\theta,-\sin \phi \sin \theta, -\cos \phi \sin\theta \right) \,, \\
& p_5^{\mu} = u_3 \, \frac{\sqrt{s}}{2}  \left(1,\cos\theta_{ll},-\sin \phi_{ll} \sin \theta_{ll}, -\cos \phi_{ll} \sin\theta_{ll} \right) \,,
\end{aligned}
\end{align}
while $p_6$ follows from momentum conservation. The values of $t$, $\cos\theta$ and $u_3$ are fixed by the constraints $p_3^2=0$, $(p_5+p_6)=M_{ll}^2$ and $p_6^2=0$, respectively.
We chose the following numerical values
\begin{align} \label{eq:unisliceparams}
s = 10^4 \, \text{GeV}^2 \,, \qquad M_{ll} = 60 \, \text{GeV}  \,, \qquad \phi = \frac{1}{10} \,, \qquad u_1 = \frac{1}{7}\,, \qquad \theta_{ll} = \frac{\pi}{2} \,, \qquad \phi_{ll} = \frac{\pi}{3} \,,
\end{align}
and generated 199 points by varying the remaining variable $u_2$ in the interval of $(87/175,29/50)$. 
In order to evaluate the hard functions we rationalise the phase-space points as described in Appendix~\ref{app:rationalise6pt}.
In Fig.~\ref{fig:hardfunctions_u2slice}, we plot the tree-level, one- and two-loop hard functions for all scattering channels 
in the $\wpaa$ and $\wmaa$ productions on the univariate phase-space slice defined above. In these figures we show both the full and the leading colour two-loop hard functions.
We additionally analyse the impact of the subleading colour contributions to the full colour hard functions. 
In Fig.~\ref{fig:hardfunctions_slc_corrections} we show the relative difference between the full colour and the leading colour hard functions, for the same set of phase-space points.
We observe that for all scattering channels in the $\wpaa$ and $\wmaa$ productions, the subleading colour corrections are negative. 
Furthermore, we found that for the $u\bar{d}\to\gamma\gamma\nu_\ell \ell^+$, $\bar{d}u\to\gamma\gamma\nu_\ell \ell^+$ and $d\bar{u}\to\gamma\gamma \ell^- \bar{\nu}_\ell$ channels,
the distributions of subleading colour corrections are sharply peaked at around $-10\%$, while for the $\bar{u}d\to\gamma\gamma \ell^- \bar{\nu}_\ell$ channel they are spread between
$-35\%$ and $-10\%$. 
We however stress that the full picture of the subleading colour corrections to the hard functions can only be drawn by using the phase-space points corresponding to a
NNLO cross section computation, with the correct kinematical cuts and renormalisation scale applied and with the infrared subtraction scheme matching the one used in the said NNLO calculation.

\begin{figure}[t!]
\begin{subfigure}{.5\textwidth}
\centering
\includegraphics[width=0.9\textwidth]{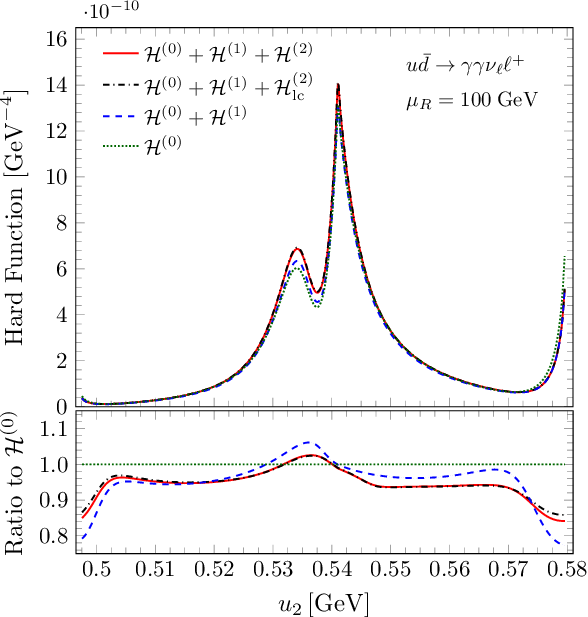}
\caption{$u\bar{d}\to\gamma\gamma\nu_\ell \ell^+$}
\end{subfigure}
\begin{subfigure}{.5\textwidth}
\centering
\includegraphics[width=0.9\textwidth]{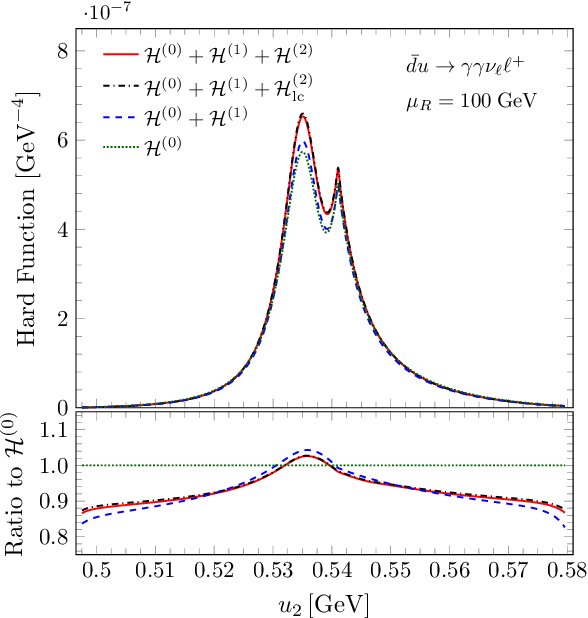}
\caption{$\bar{d}u\to\gamma\gamma\nu_\ell \ell^+$}
\end{subfigure}
\begin{subfigure}{.5\textwidth}
\centering
\includegraphics[width=0.9\textwidth]{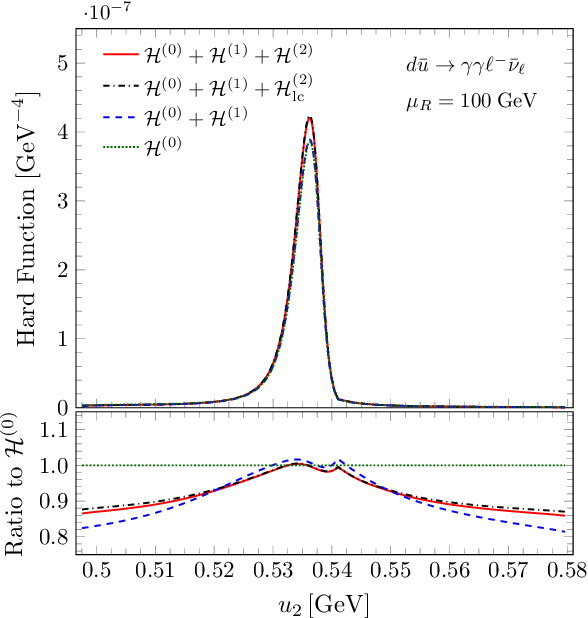}
\caption{$d\bar{u}\to\gamma\gamma \ell^- \bar{\nu}_\ell$}
\end{subfigure}
\begin{subfigure}{.5\textwidth}
\centering
\includegraphics[width=0.9\textwidth]{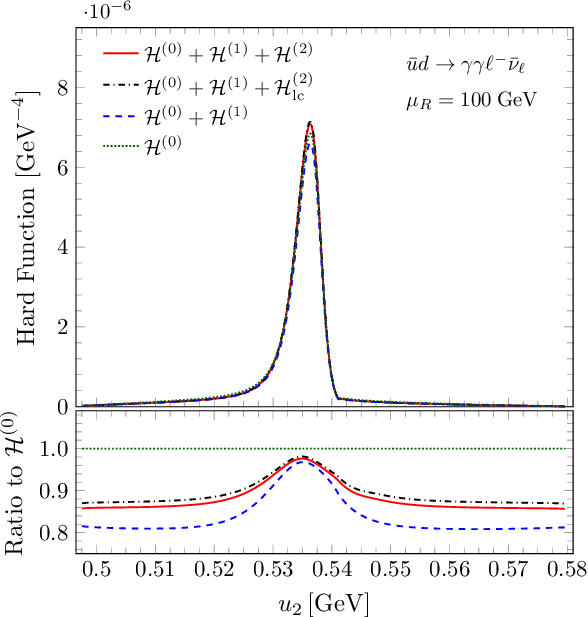}
\caption{$\bar{u}d\to\gamma\gamma \ell^- \bar{\nu}_\ell$}
\end{subfigure}
\caption{Tree level, one- and two-loop hard functions evaluated on the univariate phase-space slice defined by Eqs.~\eqref{eq:unislice1}~and~\eqref{eq:unisliceparams} 
	for all scattering channels of $\wpaa$ (upper panel) and $\wmaa$ (lower panel) production, with $\mu_R = 100$ GeV. }
\label{fig:hardfunctions_u2slice}
\end{figure}

\subsection{Analytic properties}

The results derived in this paper are the first to make use of the non-planar (one-mass) pentagon functions~\cite{Abreu:2023rco}.
Although for the subleading colour amplitudes we know the rational coefficients only numerically, this suffices to study certain analytic properties of the amplitudes related to the non-planar pentagon functions.
These properties may be conveniently exposed in terms of the \emph{letters}~\cite{Henn:2013pwa,Henn:2014qga}, algebraic functions of the kinematic invariants which dictate the singularity structure of the amplitudes.
In other words, the amplitudes may be singular only where one or more letters either vanish or go to infinity.
The pentagon functions can be written as iterated integrals with integration kernels given by logarithms of the letters~\cite{Chen:1977oja}, and this representation makes their analytic structure manifest.
The most interesting analytic properties are related to the degree-4 polynomials in the Mandelstam invariants which appear under a square root: $\Delta_5$, which is the Gram determinant of the external momenta, and $\Sigma_5^{(i)}$ with $i=1,...,6$, which are mass-deformations of $\Delta_5$ related by permutations~\cite{Abreu:2021smk}.
In particular, some non-planar Feynman integrals are non-smooth inside the physical $s_{12}$ scattering channel\textsuperscript{\ref{note-momenta}} where one of the $\Sigma_5^{(i)}$ vanishes, i.e., they are finite but their derivatives are singular. 
Furthermore, some non-planar Feynman integrals are divergent inside the physical $s_{12}$ scattering channel where $\Sigma_5^{(3)}=0$.

First of all, we observe that the letter $\sqrt{\Delta_5}$ is absent in the one- and two-loop finite remainders. 
This has by now become a well established phenomenon, as it has already been observed in a multitude of massless and one-mass two-loop five-point amplitudes, and linked ---~in the massless case~--- to cluster algebras~\cite{Chicherin:2020umh} and Gr\"obner fans~\cite{Bossinger:2022eiy}.

We further note that the set of pentagon functions denoted by ${\cal F}_{\Sigma_5}$ in Ref.~\cite{Abreu:2023rco}, that is, those involving the letters $\sqrt{\Sigma_5^{(i)}}$ (for all $i = 1,\ldots,6$), are absent from the two-loop amplitudes. 
We emphasise that the letters $\sqrt{\Sigma_5^{(i)}}$ only appear starting from two-loop order, and are therefore absent already at the level of the bare amplitudes.
This is in contrast with $\sqrt{\Delta_5}$, which instead cancels out in the subtraction of the IR and UV singularities and is therefore absent only at the level of the finite remainders.
The absence of these letters has two important consequences.
First, it implies that the amplitudes are finite inside the physical channel.
The pentagon functions which are singular at $\Sigma_5^{(3)}=0$ inside the $s_{12}$ channel in fact all contain the letter $\sqrt{\Sigma_5^{(3)}}$.
Second, evaluating numerically the set of pentagon functions ${\cal F}_{\Sigma_5}$ demands higher intermediate precision due to the presence of integrable singularities, at the cost of performance. 
Excluding these functions from the numerical evaluation of the amplitudes therefore leads to a useful speed up.
 
Finally, a number of (non-planar) pentagon functions are non-smooth when crossing the \mbox{$\Sigma_5^{(i)}=0$} hyper-surfaces inside the $s_{12}$ channel, i.e., they are finite but their derivatives are singular. 
All the weight-2 pentagon functions with this property ($F^{(2)}_{i}$ for $i = 30,\ldots,35$, one for each $\Sigma_5^{(i)}$) are present in the finite remainders.
As for the weight-3 and $4$ pentagon functions, we observe that only those which are non-smooth at \mbox{$\Sigma_5^{(3)}=0$} are present.
This observation holds also for the hard functions, as all the contributing permutations leave $\Sigma_5^{(3)}$ invariant.

The simplification of the analytic structure of the scattering amplitudes and finite remainders with respect to the Feynman integrals highlights the importance of classifying the required special functions according to their analytic properties.
With an arbitrary choice of special functions in the representation of the master integrals, in fact, the cancellations above would only occur at the numerical level.
In Refs.~\cite{Chicherin:2021dyp,Abreu:2023rco}, instead, the non-trivial analytic properties of the integrals were isolated into the minimal number of independent pentagon functions. 
This guarantees that the cancellation of, say, the divergence at $\Sigma_5^{(3)}=0$ implies the absence of the corresponding pentagon functions.
In this way the analytic structure of the amplitudes and finite remainders is more transparent, and the number of special functions which need to be evaluated is minimised.

\subsection{Description of the ancillary files}

We provide analytic expressions for the one-loop amplitudes through $\cO(\eps^2)$ and for the two-loop leading colour amplitudes through $\cO(\eps^0)$ in the ancillary files~\cite{zenodo}.
Numerical results for the subleading colour contributions are available upon request.
The analytic expressions are given separately for the finite remainders and the corresponding pole terms. 
The $L$-loop finite remainders $F^{(L)}$ truncated at order $\eps^0$ are presented in the form
\begin{equation}
	F^{(L)} = \sum_{ij} r_{i}(\vec{y}) \,  S_{ij} \, m_{j}(f) \,,
	\label{eq:finremwithsparsematrix}
\end{equation}
where $r_i$ are rational coefficients, $S$ is a sparse matrix of rational numbers, and $m_{j}(f)$ are monomials of pentagon functions, transcendental constants and square roots.
The rational coefficients $r_i$ are written in terms of independent polynomials $y_i$, functions of the five-particle momentum-twistor variables $\vec{x}$, collected globally across all finite remainders at each loop order. 
In the case of the one-loop finite remainders $F^{(1)}$ expanded up to order $\eps^2$, the sparse matrices have an additional index for the order in $\eps$, as
\begin{equation}
	F^{(1)} = \sum_{k=0}^2 \eps^k \sum_{ij} r_{i}(\vec{y}) \,  S_{kij} \, m_{j}(f) \,.
	\label{eq:finremwithsparsematrixEps2}
\end{equation}
Analytic expressions for both the poles and finite remainders are derived with $x_1$ set to 1; the rational coefficients' dependence on $x_1$ is restored in the numerical evaluation files.

All our ancillary files are in text format and contain expressions in \texttt{Wolfram Mathematica} language.
The sub-amplitudes are given in the notation of the article as follows
\begin{align}
\label{eq:subamplitudelabel}
\begin{aligned}
	\texttt{uu} & = A_{5,uu}^{(L)} \,, &
	\texttt{ud} & = A_{5,ud}^{(L)} \,, &
	\texttt{Qk} & = A_{5,q}^{(L)} \,, \\
	\texttt{uw45} & = A_{4,u}^{(L)} \,, &
	\texttt{wwaa} & = A_{3}^{(L)} \,, & 
\end{aligned}
\end{align}
while for the helicity configurations we have
\begin{align}
\label{eq:helicitylabel}
\begin{aligned}
	\texttt{pmpp1} & = \tilde{A}^{(L),+-++}_{5,uu/ud/q;1} \,, &
	\texttt{pmpp2} & = \tilde{A}^{(L),+-++}_{5,uu/ud/q;2} \,, &
	\texttt{pmpp3} & = \tilde{A}^{(L),+-++}_{5,uu/ud/q;3} \,, &
	\texttt{pmpp4} & = \tilde{A}^{(L),+-++}_{5,uu/ud/q;4} \,, \\
	\texttt{pmp1} & = \tilde{A}^{(L),+-+}_{4,u;1} \,, &
	\texttt{pmp2} & = \tilde{A}^{(L),+-+}_{4,u;2} \,, &
	\texttt{pmp3} & = \tilde{A}^{(L),+-+}_{4,u;3} \,, &
	\texttt{pmpX} & = \tilde{A}^{(L),+-+}_{4,u;4} \,, \\
	\texttt{pm1} & = \tilde{A}^{(L),+-}_{3;1} \,, &
	\texttt{pm2} & = \tilde{A}^{(L),+-}_{3;2} \,, &
	\texttt{pmA} & = \tilde{A}^{(L),+-}_{3;3} \,, &
	\texttt{pmB} & = \tilde{A}^{(L),+-}_{3;4} \,.
\end{aligned}
\end{align}
The five- and six-particle momentum-twistor variables are denoted by \verb=ex[i]=$= x_i$ ($i=1,\ldots,6$) and \verb=z[i]=$=z_i$ ($i=1,\ldots,8$), respectively, and the Mandelstam invariants by \verb=s[i,...,j]=$= (p_i+\ldots+p_j)^2$, as well as \verb=s12=$= s_{12}$, \verb=s123=$=s_{123}$, et cetera. The dimensional regulator, $\eps$, is denoted by \verb=eps=.
We list and describe below all our ancillary files.

\begin{itemize}

	\item \verb=finite_remainders/tree_waa_<X>_<helicity>.m=: tree-level amplitudes organised according to the type of sub-amplitude \verb=<X>= (see Eq.~\eqref{eq:subamplitudelabel}) and the helicity configuration \verb=<helicity>= (see Eq.~\eqref{eq:helicitylabel}). 

	\item \verb=finite_remainders/FiniteRemainder_waa_coeffs_y_<k>L_<X>_Nfp<a>_Ncp<b>_<helicity>.m=: rational coefficients ($r_i$ in Eq.~\eqref{eq:finremwithsparsematrix}). 
		The files are organised according to the loop-order \verb=<k>=, the sub-amplitude type \verb=<X>= (see Eq.~\eqref{eq:subamplitudelabel}), the powers of $n_f$ (\verb=<a>=) and $N_c$ (\verb=<b>=) and the helicity configuration \verb=<helicity>= (see Eq.~\eqref{eq:helicitylabel}).
		In order to enable the evaluation of the one-loop amplitudes up to $\cO(\eps^2)$ we have also provided the one-loop finite remainders up to $\cO(\eps^2)$. The corresponding file names are marked by the \verb=_eps2.m= ending.

	\item \verb=finite_remainders/FiniteRemainder_waa_sm_*.m=: sparse matrices of rational numbers ($S$ in Eqs.~\eqref{eq:finremwithsparsematrix} and~\eqref{eq:finremwithsparsematrix}) connecting the rational coefficients and the pentagon-function monomial basis. 
	         They are written in \texttt{Mathematica}'s \verb=SparseArray= format.
		The sparse matrix files are organised as discussed in the previous item.

	\item \verb=finite_remainders/ys_<k>L_simp.m=: replacement rules defining the common polynomial factors $\vec{y}$ (\verb=y[i]=) as functions of the five-particle momentum-twistor variables $\vec{x}$ (\verb=ex[i]=), collected across all finite remainders separately at each loop order \verb=<k>=.

	\item \verb=finite_remainders/FunctionBasis_<k>L.m=: monomial basis for the \verb=<k>=-loop finite remainders. 
	They contain pentagon functions \verb=F[w,i]=, the transcendental constants \verb=im[1, 1]= ($\i \pi$) and \verb=re[3, 1]= ($\zeta_3$), powers of $x_1$ (\verb=ex[1]=), and square roots (\verb=sqrtDelta5=, \verb=sqrtG3[i]=, \verb=sqrtSigma5[i]=).
	See Refs.~\cite{Chicherin:2021dyp,Abreu:2023rco} for more details.

	\item \verb=poles/poles_waa_<k>L_<X>_Nfp<a>_Ncp<b>_<helicity>.m=: UV and IR poles defined in Eqs.~\eqref{eq:pole1L}~and~\eqref{eq:pole2L}, organised as the finite remainders.
		In contrast to the latter, the pole terms are represented as in Eq.~\eqref{eq:bareampepsexpanded}. 
		Each files contains a list with the following format,
		\begin{center}
		\verb={coefficientrules,{coefficients,monomials}}= \,,
		\end{center}
		where \verb=coefficientrules= is a list of replacement rules defining the independent rational coefficients \verb=f[i]= as functions of the five-particle momentum-twistor variables $\vec{x}$ (\verb=ex[i]=), \verb=monomials= are pentagon-function monomials (including also square roots and transcendental constants, as discussed in the fifth item of this list), and  \verb=coefficients= are the rational coefficients of the monomials, written in terms of $\eps$ (\verb=eps=) and of the independent coefficients \verb=f[i]=. 
        
	\item \verb=decay_currents/InvDelta_*.m=: inverse of the Gram matrices defined in Eq.~\eqref{eq:GramMatrices}. They are contained in the following files:
		\begin{align}
		\begin{aligned}
			\texttt{InvDeltaM\_5pt.m} &: \quad  \Delta_5^{-1}  \,, \\
			\texttt{InvDeltaM\_4pt.m} &: \quad \Delta_4^{-1}  \,, \\
			\texttt{InvDeltaM\_4pt\_x34.m} &: \quad  \Delta_4^{-1}|_{p_3 \leftrightarrow p_4}  \,, \\
			\texttt{InvDeltaM\_3pt.m} &: \quad  \Delta_3^{-1}  \,.
		\end{aligned}
		\end{align}

	\item \verb=decay_currents/decay_*.m=: decay currents for $W\to\ell\nu$, $W\to\ell\nu\gamma$ and $W\to\ell\nu\gamma\gamma$ needed to build the full six-point amplitude from the five-, four- and three-particle
		amplitudes. The decay currents are written in terms of six-particle momentum-twistor variables $\vec{z}$ (\verb=z[i]=). 
		The following files are associated with the decay currents
		specified in Eqs.~\eqref{eq:5ptcurrentdecomposition},~\eqref{eq:4ptcurrentdecomposition4},~\eqref{eq:4ptcurrentdecomposition3}~and~\eqref{eq:3ptcurrentdecomposition}:
		\begin{align}
		\begin{aligned}
			\texttt{decay\_5pt.m} &: \quad  L_{\mu}(5_\nu,6_{\bar{\ell}}) \,, \\
			\texttt{decay\_4pt\_W4.m} &: \quad L^i_{\mu}(4_\gamma,5_\nu,6_{\bar{\ell}})  \,, \\
			\texttt{decay\_4pt\_W3.m} &: \quad L^i_{\mu}(3_\gamma,5_\nu,6_{\bar{\ell}})  \,, \\ 
			\texttt{decay\_3pt.m} &: \quad  L^i_{\mu}(3_\gamma,4_\gamma,5_\nu,6_{\bar{\ell}})  \,.
		\end{aligned}
		\end{align}
		We use the following short-hards:
                \begin{align}
                \begin{aligned}
			\texttt{InvCMWsq} & = \frac{1}{M_W^2 - \i \, \Gamma_W M_W} \,, \\
			\texttt{Wprop[s]} & = \frac{1}{s-M_W^2 + \i \, \Gamma_W M_W} \,.
		\end{aligned}
		\end{align}

	\item \verb=pentagon_functions_permutations/=: permutation rules for the one-mass pentagon functions.

	\item \verb=Evaluate_BareAmplitudes_WplusAA_RationalisedPS.wl=: \texttt{Mathematica} notebook which illustrates how to obtain the values of the leading colour bare helicity amplitudes for the $u\bar{d}\to\gamma\gamma\nu_\ell \ell^+$ scattering channel.

	\item \verb=Evaluate_HardFunctions_w2photon_RationalisedPS.wl=: \texttt{Mathematica} notebook to evaluate the leading colour hard functions for all scattering channels in the $\wpaa$ and $\wmaa$ productions. Here we rationalise the
		six-point phase-space points as described in Appendix~\ref{app:rationalise6pt}.

	\item \verb=Evaluate_HardFunctions_w2photon.wl=: \texttt{Mathematica} notebook to evaluate the leading colour hard functions for all scattering channels in the $\wpaa$ and $\wmaa$ productions without phase-space rationalisation.

	\item \verb=utilities.m=: definition of auxiliary functions required for the numerical evaluation notebooks.

\end{itemize}

The numerical evaluation scripts require the installation of the library \texttt{PentagonFunctions++}~\cite{PentagonFunctions:cpp} and its availability in \texttt{Mathematica}'s search path.

\section{Conclusion}
\label{sec:conclusions}

In this article we have presented the two-loop helicity amplitudes for the production of a $W$ boson in association with two photons at the LHC.
We derived analytic expressions for the two-loop amplitudes in the leading colour limit, including the contributions from the non-planar diagrams, whereas we provide numerical results for the subleading colour contributions.
This is therefore the first result for a two-loop $2\to3$ amplitude with an external off-shell leg involving non-planar Feynman~integrals.

The results were obtained in terms of pentagon functions~\cite{Chicherin:2021dyp,Abreu:2023rco} using a finite field framework
incorporating integrand and integral reduction techniques. 
Optimised integration-by-parts identities generated using \textsc{NeatIBP}~\cite{Wu:2023upw} led to a substantial improvement in the reduction to
master integrals. 
We employed the package \textsc{pfd-parallel}~\cite{Bendle:2021ueg} to simplify the expressions of the rational coefficients by means of multivariate partial fraction decomposition.

The basis of pentagon functions permits efficient numerical evaluation
but the algebraic complexity of the coefficients is considerable in the
subleading colour partial amplitudes. In this case it is also relevant to
consider the size of the subleading colour contributions and whether it is
beneficial to reconstruct full analytic expressions through an extremely large
number of finite field evaluations. We therefore considered an alternative
approach in which the subleading colour partial amplitudes are computed
numerically. Numerical computations modulo a prime number cannot be used
directly for evaluations of physical phase-space points and so we introduced an
alternative strategy in which four-dimensional projectors were used to ensure
the input could be given in terms of real valued, rationalised, invariants.

Using these techniques, we evaluated numerically the full colour hard functions for a univariate phase-space slice 
as a test case, and investigated the size of subleading colour corrections.
We found that, while in most cases
they are of the order of $-10\%$, for a certain channel or phase-space configuration they can be larger, reaching almost $-40\%$.
While some of those corrections are perhaps not negligible, the numerical evaluation strategy would be sufficient to provide
complete results using a re-weighting method. It should be noted however that
such statements depend on the subtraction scheme used to define the finite
remainder and the kinematic cuts applied to define the observables.

Our results also allowed us to study at amplitude level certain non-trivial analytic properties of the non-planar Feynman integrals discussed in Ref.~\cite{Abreu:2023rco}.
We observe striking cancellations and simplifications in the amplitudes with respect to the Feynman integrals.
This will enable a more efficient numerical evaluation of the amplitudes, and give useful constraints in a bootstrap context.
It would therefore be of great interest to prove that these simplifications occur for the amplitudes with these kinematics in general.

Now that all amplitude-level ingredients are available, the path is open
for combination with real radiation contributions and we look forward to phenomenological studies of this process
at the NNLO in QCD.

\section*{Acknowledgements}
We are indebted to Jakub Kry\'s for collaboration in the initial stages of this project.
We thank Rene Poncelet and Manfred Kraus for fruitful discussions,
Janko Boehm and Marcel Wittmann for assistance with \textsc{pfd-parallel},
and Dmitry Chicherin and Vasily Sotnikov for useful discussions about the analytic properties of the pentagon functions.
This project has received funding from the European Union's Horizon Europe research and innovation programme under the Marie Skłodowska-Curie grant agreement No.~101105486.
H.B.H.\ has been supported by an appointment to the JRG Program at the APCTP through the Science and Technology Promotion Fund and Lottery Fund 
of the Korean Government and by the Korean Local
Governments~--~Gyeongsangbuk-do Province and Pohang
City. S.B.~acknowledges funding from the Italian Ministry of
Universities and Research (MUR) through FARE grant R207777C4R and
through grant PRIN 2022BCXSW9. Y.Z.\ is supported from the NSF of China
through Grant No.\ 12075234, 12247103, and 12047502.
\appendix

\section{Momentum-twistor parametrisations}
\label{app:MomentumTwistors}

We define here the momentum-twistor variables which we used to obtain a rational and minimal parametrisation of the external kinematics.
For the six-particle kinematics we adopt the parametrisation of Ref.~\cite{Hartanto:2019uvl},
\begin{align}
\begin{aligned}
  z_1 &=  s_{12} \,, &
  z_2 &= -\frac{\trp(1234)}{s_{12}s_{34}} \,, &
  z_3 &= -\frac{\trp(1345)}{s_{45}s_{13}} \,,  \\
  z_4 &= -\frac{\trp(1456)}{s_{14}s_{56}} \,, &
  z_5 &=  \frac{s_{23}}{s_{12}} \,, &
  z_6 &= -\frac{\trp[15(3+4)2]}{s_{12}s_{15}} \,,  \\
  z_7 &=  \frac{\trp[51(2+3)(2+3+4)]}{s_{15} s_{23}} \,, & 
  z_8 &=  \frac{s_{123}}{s_{12}} \,, &
\end{aligned}
\label{eq:momtwistor6pt}
\end{align}
where
\begin{align}
& \tr_{\pm}(ij \cdots kl) = \frac{1}{2} \tr\left[(1\pm\gamma_5)\slashed{p}_i\slashed{p}_j \cdots\slashed{p}_k\slashed{p}_l\right] \,, \\
& \tr_{\pm}[\cdots (i+j) \cdots] = \tr_{\pm}(\cdots i \cdots) + \tr_{\pm}(\cdots j \cdots) \,. 
\end{align}
The five-, four- and three-particle $W$-production amplitudes can instead be described through a five-particle parametrisation.
We use that of Ref.~\cite{Badger:2021ega},
\begin{equation}
\begin{alignedat}{2}
  & x_1 = s_{12} \,, \qquad && x_2 = -\frac{\trp(1234)}{s_{12}s_{34}} \,, \\
  & x_3 = \frac{\trp[1341(5+6)2]}{s_{13}\trp{[14(5+6)2]}} \,, \qquad && x_4 = \frac{s_{23}}{s_{12}} \,, \\
  & x_5 = -\frac{\trm[1(2+3)(1+5+6)(5+6)23]}{s_{23}\trm[1(5+6)23]} \,, \qquad && x_6 = \frac{s_{123}}{s_{12}} \,.
\end{alignedat}
\label{eq:momtwistor5pt}
\end{equation}
Note that the spinor-phase information is lost when using momentum-twistor variables.
Care should therefore be taken to restore it in order to perform operations such as permutations and parity conjugation, 
e.g.\ by introducing suitable helicity-dependent spinor-phase factors.
We refer to Appendix~C of Ref.~\cite{Badger:2023mgf} for a discussion of this topic.

\section{Rationalising the six-particle phase-space points}
\label{app:rationalise6pt}

The evaluation of the rational coefficients of the pentagon-function monomials in the two-loop five-particle subleading colour amplitudes is done 
numerically over finite field as detailed in 
Section~\ref{sec:numericalframework}. For this approach, we need the Mandelstam invariants that parametrise the 
five-particle amplitude to be rational numbers as we are computing the unreduced contracted amplitudes $T^\dagger_j \cdot \tilde{A}^{(L)}_{5,uu/ud,k}$ 
(c.f.\ Eq.~\eqref{eq:contractedampsij}). In addition, rationalising the six-particle momentum-twistor variables $z_i$ does not guarantee that the resulting Mandelstam invariants are real valued, and this is required for the evaluation of the pentagon functions. 
Here we describe our procedure to obtain rationalised Mandelstam invariants.

	 We start with a set of six-particle momenta, defined as all outgoing, following Eqs.~\eqref{eq:processdefinition}~and~\eqref{eq:momentumconservation}. 
	      We then compute 8 independent six-particle Mandelstam invariants, which we choose as
         \begin{equation}
                \vec{s}_6=\lbrace s_{12}, s_{23}, s_{34}, s_{45}, s_{56}, s_{16}, s_{123}, s_{234} \rbrace \,,
               \label{eq:sij6pt}
         \end{equation}
         and rationalise them to the target accuracy. 
         Increasing the accuracy in the rationalisation results in larger rational numbers, 
	 which means that more prime fields are needed for the rational reconstruction. 
	 The evaluation time of the rational coefficients in this approach scales linearly with the number of required primes,\footnote{This process is sequential and stops when including a new prime does not change the reconstructed number.} which should therefore be minimised.
	 Only six of the six-particle Mandelstam invariants contribute to the five-particle sub-amplitudes:
         \begin{equation}
                \vec{s}_5=\lbrace s_{12}, s_{23}, s_{34}, s_{56}, s_{123}, s_{234} \rbrace \,.
               \label{eq:sij5pt}
         \end{equation}
	In order to fully specify the six-particle kinematics, we also need a pseudo-scalar quantity, which captures the parity degree of freedom. We choose it to be $\tr_5(1234)$, with
         \begin{align}
		 \tr_5(ijkl) = 4 \, \i \, \varepsilon_{\mu\nu\rho\sigma} \, p_i^{\mu} p_j^{\nu} p_k^{\rho} p_l^{\sigma} \,,
		 \label{eq:tr5definition4vec}
         \end{align}
	where $\varepsilon_{\mu\nu\rho\sigma}$ is the anti-symmetric Levi-Civita pseudo-tensor.
	Since we have rationalised the Mandelstam invariants $\vec{s}_6$, we compute $\tr_5(1234)$ directly from them.
	To this end, note that the square of $\tr_5(1234)$ equals the Gram determinant of the momenta in its argument,
	\begin{align}
	\begin{aligned}
		& G_5 = \det(2p_i \cdot p_j) \bigl|_{i,j=1,\dots,4} \,, \\
		    & \phantom{G_5} = s_{12}^2(s_{23} - s_{234})^2 + (s_{123}(s_{234} - s_{34}) + s_{23}(s_{34} - s_{56}))^2  \\
		    &  \phantom{G_5 = } + 2 s_{12} \big[ s_{123}(s_{23} - s_{234}) s_{234} + s_{123}(s_{23} + s_{234})s_{34}  \\ 
		    & \phantom{G_5 = + 2 s_{12} \big[ } + s_{23}(-2 s_{34}s_{56} - s_{23}(s_{34} + s_{56}) + s_{234}(s_{34} + s_{56})) \big] \,,
	\end{aligned}
	\end{align}
	which is a function of the $\vec{s}_6$.
	For real momenta, $\tr_5(1234)$ is purely imaginary and $G_5$ is negative.
        We can therefore express $\tr_5(1234)$ in terms of $\vec{s}$ up to its sign, which we fix from the four vectors in Eq.~\ref{eq:tr5definition4vec}.
        Explicitly, we have that
	\begin{equation}
		\tr_5(1234) = \mathrm{sign}\left[  \varepsilon_{\mu\nu\rho\sigma} \, p_i^{\mu} p_j^{\nu} p_k^{\rho} p_l^{\sigma}  \right] \, \i \, \sqrt{-G_5} \,.
		 \label{eq:tr5definitionGram}
	\end{equation}
        The values of all other pseudo-scalar invariants $\tr_5(ijkl)$ can be obtained from $\tr_5(1234)$ and $\vec{s}_6$, as the product $\tr_5(1234) \, \tr_5(ijkl)$ can be expressed in terms of $\vec{s}_6$ through Gram determinants.
       We provide in the ancillary files the expression of $\tr_5(1234) \tr_5(ijkl)$ in terms of $\vec{s}_6$ for all the required $\tr_5(ijkl)$.
   
	Once we have the rationalised six- and five-particle invariants, we compute the six- and five-particle momentum-twistor variables, $z_i$ and $x_i$ respectively, from them. 
	To this end, we first rewrite the momentum-twistor variables defined in Appendix~\ref{app:MomentumTwistors} in terms of $\vec{s}_6$ and pseudo-scalar invariants $\tr_5(ijkl)$.
	For the six-particle momentum-twistor variables we have
	\begin{align}
	\begin{aligned}
	\label{eq:mtzdefinition}
	& z_1 = s_{12} \,, \\
	& z_2 = \frac{s_{12}s_{23}-s_{12}s_{234} + s_{123} s_{234} - s_{123} s_{34} + s_{23} s_{34} - s_{23} s_{56} - \tr_5(1234)}{2 s_{12} s_{34}} \,, \\
	& z_3 = - \frac{(s_{16} - s_{23}) s_{34} + s_{123} (s_{34} - s_{345}) + (s_{23} - s_{234}) s_{345} + (s_{12} + s_{234}) s_{45} + (s_{345} - s_{45}) s_{56}}{2(s_{12}-s_{123} + s_{23}) s_{45}} \\
            & \phantom{z_3 = } + \frac{\tr_5(1345)}{2(s_{12}-s_{123} + s_{23}) s_{45}} \,, \\
	& z_4 = \frac{s_{123} (s_{234}-s_{16} ) - s_{234} s_{45} + (s_{16} - s_{23} + s_{45}) s_{56} - \tr_5(1456)}{2 s_{56} (s_{23} - s_{234} + s_{56}-s_{123})} \,, \\
	& z_5 = \frac{s_{23}}{s_{12}} \,, \\
	& z_6 = \frac{s_{12} s_{16} - s_{12} s_{234} + s_{16} s_{34} - s_{234} s_{345} - s_{16} s_{56} + s_{345} s_{56} - \tr_5(1256)}{2 s_{12} (s_{16} - s_{234} + s_{56})} \,, \\
	& z_7 = \frac{2 s_{16} s_{23} - 2 s_{23} s_{234} + s_{123} (s_{234}-s_{16}) +  s_{234} s_{45} + (s_{16} + s_{23} - s_{45}) s_{56} - \tr_5(1456)}{2 s_{23} (s_{16} - s_{234} + s_{56})} \,, \\
	& z_8 = \frac{s_{123}}{s_{12}} \,,
	\end{aligned}
	\end{align}
   	while for the five-particle ones we have
	\begin{align}
	\begin{aligned}
	\label{eq:mtxdefinition}
		& x_1 = s_{12} \,, \\
		& x_2 = \frac{s_{12} (s_{23} - s_{234}) + s_{123} s_{234} + (s_{23}-s_{123}) s_{34} - s_{23} s_{56}  - \tr_5(1234)}{2 s_{12} s_{34}}  \,, \\
		& x_3 = \frac{(s_{12} + s_{234} - s_{34}) \left[ s_{12} (s_{234}-s_{23}) + s_{23} s_{34} - s_{123} (s_{234} + s_{34})\right]}{
			    2 (s_{12} - s_{123} + s_{23}) \left[ s_{123} (s_{12} + s_{234} - s_{34}) - (s_{12} + s_{23}) s_{56} \right] } \\
		    & \phantom{x_3 =} + \frac{s_{56} \left[(s_{123} + s_{23}) s_{234} - s_{12} s_{234} - s_{123} s_{34} + 2 s_{12} (s_{23} + s_{34}) - s_{23} s_{56} \right]}{
			            2 (s_{12} - s_{123} + s_{23}) \left[ s_{123} (s_{12} + s_{234} - s_{34}) - (s_{12} + s_{23}) s_{56} \right]} \\
			    & \phantom{x_3 =} + \frac{(s_{34} + s_{56} -s_{12} - s_{234} ) \tr_5(1234)}{2 (s_{12} - s_{123} + s_{23}) \left[ s_{123} (s_{12} + s_{234} - s_{34}) - (s_{12} + s_{23}) s_{56} \right]} \,, \\
		& x_4 = \frac{s_{23}}{s_{12}}\,, \\
		& x_5 = \frac{ \left[ s_{12} (s_{23} + s_{234}) + s_{123} (s_{234} - s_{34}) + s_{23} s_{34} \right] s_{56}}{2 s_{23} \left[ s_{12} s_{234} + (s_{234} - s_{34}) (s_{234} - s_{56}) \right]} \\
		    & \phantom{x_5 =} + \frac{2 (s_{123} - s_{23}) s_{234} (s_{12} + s_{234} - s_{34}) + s_{23} s_{56}^2}{2 s_{23} \left[(s_{234} - s_{34}) s_{56} -s_{234} (s_{12} + s_{234} - s_{34}) \right]} \\
		    & \phantom{x_5 =} - \frac{s_{56} \tr_5(1234)}{2 s_{23} \left[ s_{12} s_{234} + (s_{234} - s_{34}) (s_{234} - s_{56}) \right]} \,, \\
		& x_6 = \frac{s_{123}}{s_{12}}\,. \\
	\end{aligned}
	\end{align}
	
We note that the expression of the six-particle momentum-twistor variables $z_i$ in Eq.~\eqref{eq:mtzdefinition} contains a Mandelstam invariant, $s_{345}$, that is not contained in $\vec{s}_6$. It can be written in terms of $\vec{s}_6$ through the vanishing of the six-particle Gram determinant,
\begin{equation}
	G_6 = \det(2p_i\cdot p_j)\bigl|_{i,j=1,\dots,5} = 0 \,,
\end{equation}
which holds for four-dimensional momenta $p_i$. 
This constraint has two solutions for $s_{345}$. 
We choose the one which matches the value of $s_{345}$ obtained directly from the floating-point momenta.
Furthermore, due to the pseudo-scalar invariants $\tr_5(ijkl)$ that appear in the definition of the momentum-twistor variables in Eqs.~\eqref{eq:mtzdefinition}~and~\eqref{eq:mtxdefinition}, $z_i$ and $x_i$
are no longer rational in general. However, we only need $\vec{s}_5$ to be rational, 
so at this stage we evaluate the momentum-twistor variables as floating-point numbers.

\bibliographystyle{JHEP}
\bibliography{waa_2L}

\end{document}